\documentclass[
    aps,
    prd,
    notitlepage,
    twocolumn,s
    tightenlines,
    nofootinbib,
    superscriptaddress]{revtex4-2}
\pdfoutput=0
\usepackage{graphicx,amssymb,amsmath,amsthm,amsfonts,epsfig,epsf}
\usepackage[linktocpage]{hyperref}
\usepackage[usenames]{color}
\usepackage{epstopdf}
\usepackage{bm}
\usepackage{dcolumn}
\usepackage[latin1]{inputenc}
\usepackage{latexsym}
\usepackage{rotating}
\usepackage{hyperref}
\usepackage{color}
\usepackage{longtable}

\usepackage{enumerate}
\usepackage{tensor,multirow}
\usepackage{url}
\usepackage{tikz,tikz-3dplot}
\usepackage{slashed}
\usepackage{ulem}
\begin{document}
\title{Waveform accuracy and systematic uncertainties \\ in current gravitational wave observations}% Force line breaks with \\%
\author{Caroline B. Owen }
\email{cbo4@illinois.edu}
\affiliation{%
Illinois Center for Advanced Studies of the Universe, Department of Physics,
University of Illinois at Urbana-Champaign, Urbana, IL 61801, USA}%
\author{Carl-Johan Haster}
\affiliation{Department of Physics and Astronomy, University of Nevada, Las Vegas, 4505 South Maryland Parkway, Las Vegas, NV 89154, USA}
\affiliation{Nevada Center for Astrophysics, University of Nevada, Las Vegas, NV 89154, USA}
\author{Scott Perkins}
\email{perkins35@llnl.gov}
\affiliation{%
Illinois Center for Advanced Studies of the Universe, Department of Physics,
University of Illinois at Urbana-Champaign, Urbana, IL 61801, USA}%
\author{Neil J. Cornish}
\affiliation{eXtreme Gravity Institute, Department of Physics, Montana State University, Bozeman, MT 59717, USA}
\author{Nicol\'as Yunes}
\affiliation{%
Illinois Center for Advanced Studies of the Universe, Department of Physics,
University of Illinois at Urbana-Champaign, Urbana, IL 61801, USA}%
\date{\today}

%%%%%%%%%%%%%%%%%%%%%%%%%%%%%%%%%%%%%%%%%%%%%%%%%%%%%%%%%%%%%%%%%%%%%%%%%%%%%%%%%%%%%%%%%%%%%%%%%
\begin{abstract}
The post-Newtonian formalism plays an integral role in the models used to extract information from gravitational wave data, but models that incorporate this formalism are inherently approximations.
Disagreement between an approximate model and nature will produce mismodeling biases in the parameters inferred from data, introducing systematic error. 
We here carry out a \textit{proof-of-principle} study of such systematic error by considering signals produced by quasi-circular, inspiraling black hole binaries through an injection and recovery campaign. 
In particular, we study how unknown, but calibrated, higher-order post-Newtonian corrections to the gravitational wave phase impact systematic error in recovered parameters.
As a first study, we produce injected data of non-spinning binaries as detected by a current, second-generation network of ground-based observatories and recover them with models of varying PN order in the phase.
We find that the truncation of higher order ($>$3.5) post-Newtonian corrections to the phase can produce significant systematic error even at signal-to-noise ratios of current detector networks. 
We propose a method to mitigate systematic error by marginalizing over our ignorance in the waveform through the inclusion of higher-order post-Newtonian coefficients as new model parameters.
We show that this method can reduce systematic error greatly at the cost of increasing statistical error. 

\end{abstract}
%%%%%%%%%%%%%%%%%%%%%%%%%%%%%%%%%%%%%%%%%%%%%%%%%%%%%%%%%%%%%%%%%%%%%%%%%%%%%%%%%%%%%%%%%%%%%%%%%
\maketitle
\section{\label{SEC:Intro}Introduction}
Gravitational waves emitted during the coalescence of compact-object binaries offer a unique way to directly observe strong-gravity systems. 
Models of the signals produced by coalescence events, derived from general relativity or modified theories of gravity, are used to extract information about the astrophysical sources from the gravitational wave data. 
However, due to the complexity of the theories and computational time constraints, these models are necessarily approximations (either numerical, analytical, or some hybrid of the two). 
Mismatch between an approximate model and nature will always result in biases in the parameters inferred from data. 
Such bias is known as mismodeling systematic error.  

In this study, we investigate mismodelling error in the context of the post-Newtonian (PN) approximation to binary inspiral signals. While contemporary analyses use more sophisticated models that include the merger and ringdown phase of the signal, we use the simpler PN inspiral regime to illustrate the effects of mismodeling and how these effects can be mitigated. Similar conclusions will apply when using full inspiral-merger-ringdown waveforms.

\begin{figure}[t]
\centering
\includegraphics[width=.8\linewidth]{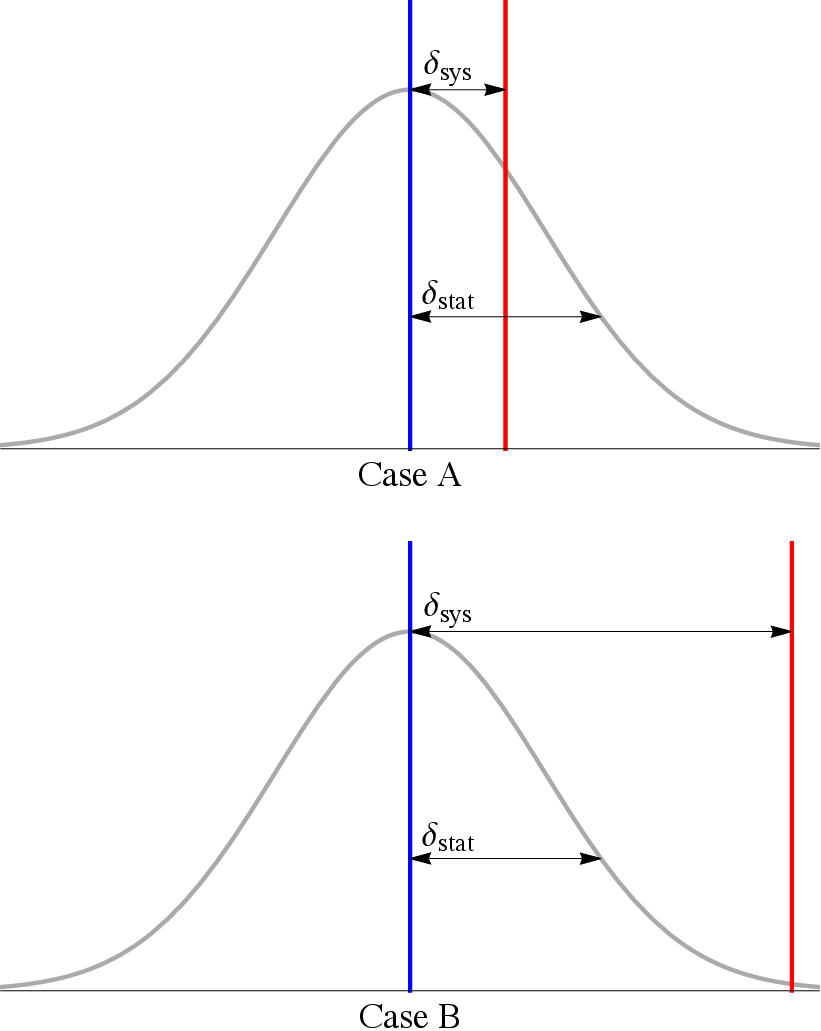} 
\caption{\label{FIG:errorfig} 
Schematic diagram to illustrate the relationship between systematic error $\delta_\mathrm{sys}$ and statistical error $\delta_\mathrm{stat}$.  
The gray curve in each panel depicts a one-dimensional, inferred posterior probability distribution on a hypothetical parameter obtained by analyzing a synthetic noise-free data set and recovering it with some model that has some amount of inherent inaccuracy with respect to the true signal. 
The injected (true) value of the parameter is indicated with a red line, while the recovered value, taken in this case from the maximum posterior point, is indicated with a blue line. 
In Case A, the systematic error is smaller than the statistical error and the value of the injected parameter lies within the credible intervals of the recovered best-fit value of the parameter. 
In Case B, however, the systematic error is dominant and the injected value lies outside of the credible interval.}
\end{figure}

The approximate nature of waveform models means that some amount of systematic error is unavoidable.
This systematic error is tolerable so long as it is significantly smaller than the statistical error caused by the finite signal-to-noise ratio (SNR) of the observations.
Figure~\ref{FIG:errorfig} shows a schematic diagram to illustrate why this is the case. 
Both panels in the figure represent a one-dimensional posterior probability distribution on a hypothetical parameter obtained by analyzing some synthetic noise-free data using some model. 
The locations of the injected (true) and recovered values for that parameter are indicated, where the recovered value is taken from the maximum posterior point.
In this idealized context, the systematic error is given by the difference between the recovered and injected values of the parameter. In data with noise, the offset will be a combination of systematic and statistical error.
The statistical error is reflected by the width of the posterior probability distribution and is due to the noise weighting in the likelihood.

In Case A, the systematic error is smaller than the statistical error, and therefore, the injected parameter is within the relevant credible interval of the recovered parameter. 
In Case B, however, the systematic error is dominant and the true value lies outside of the credible interval. 

One strives to create models that are accurate enough such that the posterior probability distributions on every parameter are reliable (i.e.~they resemble Case A in Fig.~\ref{FIG:errorfig}).
However, statistical error scales inversely with the SNR while systematic error is independent of the SNR.
Therefore, as instruments are upgraded and improved, the systematic error can become dominant. 
As we progress in our capacity to observe gravitational waves, understanding how mismodeling impacts parameter estimation becomes ever more important.

The importance of bias from mismodeling depends on the accuracy of the models used, but the construction of models is a complicated matter due to the intrinsic non-linearity of the theory and the broad parameter space of interest to us.
On solar system scales, Newtonian gravity is adequate for most calculations. 
However, in the extreme environments that produce the gravitational waves observable by ground-based detectors, Newtonian gravity does not suffice. 
Instead, one must account for post-Newtonian corrections to Newtonian gravity. 
These corrections are derived through a formal expansion of the field equations (Einstein's or otherwise) in powers of $(v/c)^2$~\cite{Blanchet:2002av}, where $v$ is the characteristic speed of the system and $c$ is the speed of light and gravity.

During the quasi-circular inspiral of compact objects, a waveform based on the PN approximation has been shown to be highly accurate when enough terms are kept in the small-velocity expansion~\cite{Will:2011nz}, but this is not the case for sufficiently massive binaries.  
As the total mass of the system increases, the merger and post-merger parts of the coalescence signal become more dominant for current ground-based detectors, and the classic pre-merger inspiral PN expansion is no longer sufficient. 
This fact has led to the creation of two classes of inspiral-merger-ringdown (IMR) models: effective-one-body (EOB) waveforms~\cite{Buonanno:1998gg, Bohe:2016gbl,Cotesta:2018fcv, Nagar:2018zoe,Nagar:2020pcj, Ossokine:2020kjp} and phenomenological waveforms~\cite{Husa:2015iqa, Khan:2015jqa, Pratten:2020fqn, Pratten:2020ceb}.

In this paper, we consider the phenomenological framework, which defines the waveform as a piecewise function, with one piece modeling the inspiral, one piece the late inspiral and merger, and one piece the post-merger.
In the phenomenological waveform model {\tt{IMRPhenomD} }\cite{Husa:2015iqa,Khan:2015jqa}, the inspiral piece is constructed from the classic 3.5PN order approximation\footnote{A term of order $(v/c)^{2n}$ relative to the leading-order term is considered to be of $n$PN order.}, enhanced with 4, 4.5, 5 and 5.5PN terms that are calibrated to a suite of EOB and numerical relativity simulations. 
The late inspiral and merger piece and the post-merger piece are modeled in a similar way, by fitting coefficients in an ansatz function to numerical relativity simulations when needed. 
Both the EOB and the phenomenological waveforms have been validated by showing high matches against a set of numerical relativity simulations~\cite{Hinder:2013oqa} and are used today for gravitational wave parameter estimation by the LIGO/Virgo collaboration~\cite{LIGOScientific:2014pky,TheVirgo:2014hva,LIGOScientific:2021usb, LIGOScientific:2021djp}. 

The post-Newtonian approximation plays a critical role in the creation of waveforms
\footnote{One exception to this is the family of Numerical Relativity Surrogate waveform models, constructed directly from catalogs of numerical relativity waveforms~\cite{Varma:2019csw, Islam:2022laz}.} either because it is later resummed as it is in EOB waveforms or because it is enhanced with fitting coefficients as it is in both the phenomenological and EOB models.
However, the inherent approximation will necessarily introduce systematic uncertainties, and therefore, systematic bias has been studied in depth over the last three decades~\cite{Cutler:2007mi, Canitrot:2001hc,
Moore:2021eok,
Littenberg:2012uj, Purrer:2019jcp,Chua:2019wgs}.

Early studies of mismodeling relied mostly on the Fisher information, an approximation to the full likelihood that is only accurate for sufficiently loud signals in Gaussian noise~\cite{Vallisneri:2007ev}. 
Cutler and Vallisneri \cite{Cutler:2007mi} developed a Fisher information formalism to estimate mismodeling bias and applied it to a 3.5PN inspiral signal recovered with a 3PN inspiral model. 
Focusing on binary black hole signals detected by LISA \cite{amaroseoane2017laser,Kaiser:2020tlg} with SNR of 1000, they concluded that it is possible for systematic error to dominate the statistical error by several orders of magnitude for intrinsic parameters like masses and spins. 
This paper was preceded by the work of Canitrot \cite{Canitrot:2001hc}, who used match filtering techniques to show that recovering a 2.5PN binary black hole signal detected by Virgo \cite{TheVirgo:2014hva} with a 2PN model can produce systematic error of order $\sim 1\%$ in the chirp mass and order $\sim 50\%$ in the symmetric mass ratio.

Moore et al. \cite{Moore:2021eok} used the formalism developed by Cutler and Valisneri \cite{Cutler:2007mi} to understand how waveform inaccuracies could lead to evidence for deviations from general relativity even when the observed signal is described by general relativity. Particularly, they considered how such evidence can accumulate when considering large catalogs of data. The authors determined that false deviations of general relativity could be detected in catalogs of 10-30 events with SNRs above 20.

Fisher studies, however, are not necessarily accurate for low SNR events, like those detected by current ground-based detectors, and thus, a Bayesian approach is preferred. 
To this end, recent studies have employed injection and recovery campaigns: synthetic data (with a set of known injection parameters) is generated using some model, and then the data is analyzed with another model. 
Through this approach, one can then determine whether systematic bias is a concern by comparing the injected parameters to the posterior probability distributions obtained.

Littenberg et al. performed such an analysis, injecting numerical relativity waveforms and recovering with EOB models  \cite{Littenberg:2012uj}. 
Focusing on stellar-mass binary black holes observed by advanced LIGO \cite{LIGOScientific:2014pky} and Virgo\cite{TheVirgo:2014hva}, they found that at SNRs $<$ 50 systematic error was smaller than or comparable to statistical error for a wide range of mass ratios. 
However, for SNRs $>$ 100, as is expected from 3G detectors, systematic error can dominate. 
More recently, P\"urrer and Haster also injected numerical relativity waveforms and recovered with various semi-analytic frequency-domain models \cite{Purrer:2019jcp}.
Focusing on 3G events at SNRs $\in [466, 2598]$ that spend much longer in the 3G sensitivity band than those studied in~\cite{Littenberg:2012uj}, the authors concluded that systematic error induced by these semi-analytic models needs to be reduced by three orders of magnitude.

Evidence already exists from observations during the O3 LIGO/Virgo campaign that systematic error may be influencing posterior probability distributions~\cite{LIGOScientific:2021usb, LIGOScientific:2021djp}. 
Indeed, for certain O3 events, the LIGO/Virgo collaboration found somewhat different posterior probability distributions for various source parameters when analyzing the data with the {\tt IMRPhenomXPHM}~\cite{Pratten:2020ceb} or {\tt SEOBNRv4PHM}~\cite{Ossokine:2020kjp} models.
The analyses of  GW191219\_163120 with these two models produced differences in the inferred spins and mass ratios, analyses of GW191109\_01071 produced differences in the inferred inclination angle, total mass and distance, analyses of  GW200129\_065458 produced differences in the evidence for precession and the inferred mass ratio, and analyses of GW200208\_222617 produced a multimodal mass posterior probability distribution with different waveforms preferring difference modes \cite{LIGOScientific:2021djp}.

In these analyses of O3 observations, different data analysis pipelines were used when performing parameter estimation with the IMRPhenomXPHM and  SEOBNRv4PHM models \cite{LIGOScientific:2021usb, LIGOScientific:2021djp}. Differences in sampling methods, rather than discrepancies between models, might have been partly responsible for some of the biases seen. However, other waveform systematic studies that used the same sampling methods still found biases in recovered parameters, albeit using different waveform models\cite{LIGOScientific:2020ufj,LIGOScientific:2016ebw,Islam:2020reh}. When taken together, these examples are evidence that a better
understanding of systematic error is necessary.

 In spite of all of the work in the study of systematic errors, no work has yet been done to employ Bayesian methods to assess the inaccuracies of the latest waveform models with respect to our ignorance of (unknown) high-order PN order terms in the era of current ground-based detectors.
In this paper we embark on a \textit{proof-of-principle} study to determine whether the unknown, yet numerically-calibrated, PN terms in the {\tt IMRPhenomD} \cite{Husa:2015iqa,Khan:2015jqa} waveform model can systematically bias the extraction of intrinsic source parameters (like the masses) at SNRs consistent with current and near-future observations with LIGO/Virgo/KAGRA \cite{KAGRA:2013rdx}.
We focus on inspiraling black hole binaries of mass ratios that are detected by a current-generation gravitational wave detector network at SNRs of ${\cal{O}}(10)$. 
We consider an \textit{inspiral-only} {\tt IMRPhenomD} model created by stopping the {\tt IMRPhenomD} waveform at the transition frequency between the inspiral and intermediate regions of the waveform. 
We produce injected data from this model and recover the injected data using models inspired by the injection model but truncated at 5, 4.5, 4, or 3.5PN order in the inspiral phase.  
We choose to work with {\tt IMRPhenomD} because its structure makes it straightforward to implement a study such as the one described above. Conclusions drawn in this study provide no information about the accuracy of {\tt IMRPhenomD} when analyzing actual gravitational wave data. Rather, this study allows us to quantify the relative importance to parameter estimation of the higher-order PN terms. We expect one would obtain similar results using any waveform model based on the inspiral PN formalism, although we do not formally show this here.

Using standard Bayesian inference packages, we then explore the parameter space of interest to construct posterior probability distributions.
From these distributions, we can then estimate the statistical and systematic errors in the inferred parameters.

Overall, we find that the truncation of the {\tt IMRPhenom} family even at 5PN order can lead to systematic biases larger than statistical uncertainties in the chirp mass, mass ratio and individual masses already at the SNRs expected in the fourth observing run of the LIGO/Virgo/KAGRA detectors. 
Of course, as stated above, this does not necessarily imply that the {\tt IMRPhenomD} family contains these intrinsic errors since the 4, 4.5, 5, and 5.5PN terms are never set to zero in any of the {\tt IMRPhenomD} models used in the analysis of real astrophysical observations. 
Our results, however, do imply that fitting errors in the 4, 4.5, 5, and 5.5PN coefficients as well as the absence of higher order, currently-unknown PN terms could introduce systematic errors that are not currently being accounted for. 
These fitting errors arise due to both the finite and discrete nature of the numerical relativity simulation sets used in the fit, as well as possibly the choice of the fitting functions. 

In order to deal with these potential systematic uncertainties, we propose a method to ameliorate them: to include our ignorance of the missing terms directly in the model and then to marginalize over them. 
More specifically, we propose that uncertainties be included in the waveform model as higher PN order parameters, with physically uninformed (uniform) priors that encapsulate the residual from the fits. 
We then propose to marginalize over these systematic-uncertainty parameters simultaneously while exploring the astrophysically informative parameter space. 
We show that the effect of the inclusion of these new systematic parameters is to significantly reduce the bias (by pushing the peak of the posterior probability distribution back to the injected value), at the cost of increasing the statistical error (by widening the distribution).  

The remainder of this paper is structured as follows: in Sec. \ref{SEC:The Model} we review the waveform models used in our analyses, followed by Sec. \ref{SEC:ExperimentalDesign} where we lay out the details of the experimental design of our injection and recovery campaign.
The results of this campaign are presented in Sec. \ref{SEC:Results}. 
In Sec. \ref{SEC:MissingTerms}, we explore a way to account for our ignorance of the true waveforms by marginalizing over missing terms in our model. 
And in Sec. \ref{SEC:Discussion} we provide discussion on the results obtained in this study.
In what follows, we will work in geometric units where $G=1=c$.
%

%%%%%%%%%%%%%%%%%%%%%%%%%%%%%%%%%%%%%%%%%%%%%%%%%%%%%%%%%%%%%%%%%%%%%%%%%%%%%%%%%%%%%%%%%%%%%%%%%
\section{A Truncated IMRPhenomD Model to Represent Mismodeling}\label{SEC:The Model}

In this section, we provide a summary of the {\tt IMRPhenomD} waveform approximant, which is a phenomenological frequency domain waveform model for spin-aligned black hole binaries \cite{Husa:2015iqa,Khan:2015jqa}. In doing so, we generally follow the notation of \cite{Husa:2015iqa,Khan:2015jqa}. 
The frequency domain strain is a complex function, expressed in polar form as 
\begin{align}
    \tilde{h}(f,\bm{\theta}) = A(f,\bm{\theta})e^{-i\phi(f,\bm{\theta})},
\end{align}
where $f$ is the frequency of the gravitational wave and $\bm{\theta}$ is a vector of waveform parameters. 
This approximant includes only the dominant quadrupolar radiation mode.  
The phenomenological modeling of the strain is broken up into three regions: the inspiral at low frequencies, the post-merger and ringdown at high frequencies, and an intermediate region for the plunge and merger.%

In this proof-of-principle study, we will make some simplifications to render the problem tractable.
First, we focus on the inspiral region alone, as this is where the PN formalism is valid.
This simplification limits our analysis to relatively low (total) mass sources, such that the total SNR from the whole coalescence is dominated by the inspiral region.
Second, we focus only on the accuracy of the phase of the inspiral frequency-domain strain $\phi(f,\bm{\theta})$ and leave the amplitude as is.
We do so because the phase of the gravitational wave signal provides the most constraining information for the inference of binary parameters.
In order to minimize the effects of both amplitude modulations and subdominant waveform harmonics, we limit our study to quasi-circular binary inspirals, with black hole spins aligned or anti-aligned with the orbital angular momentum and with comparable mass ratios.
In spite of these approximations, the study we carry out here will still allow conclusions that should apply generally to a large number of gravitational wave sources.
\begin{figure}[t]
  \centering
  \includegraphics[width=\linewidth]{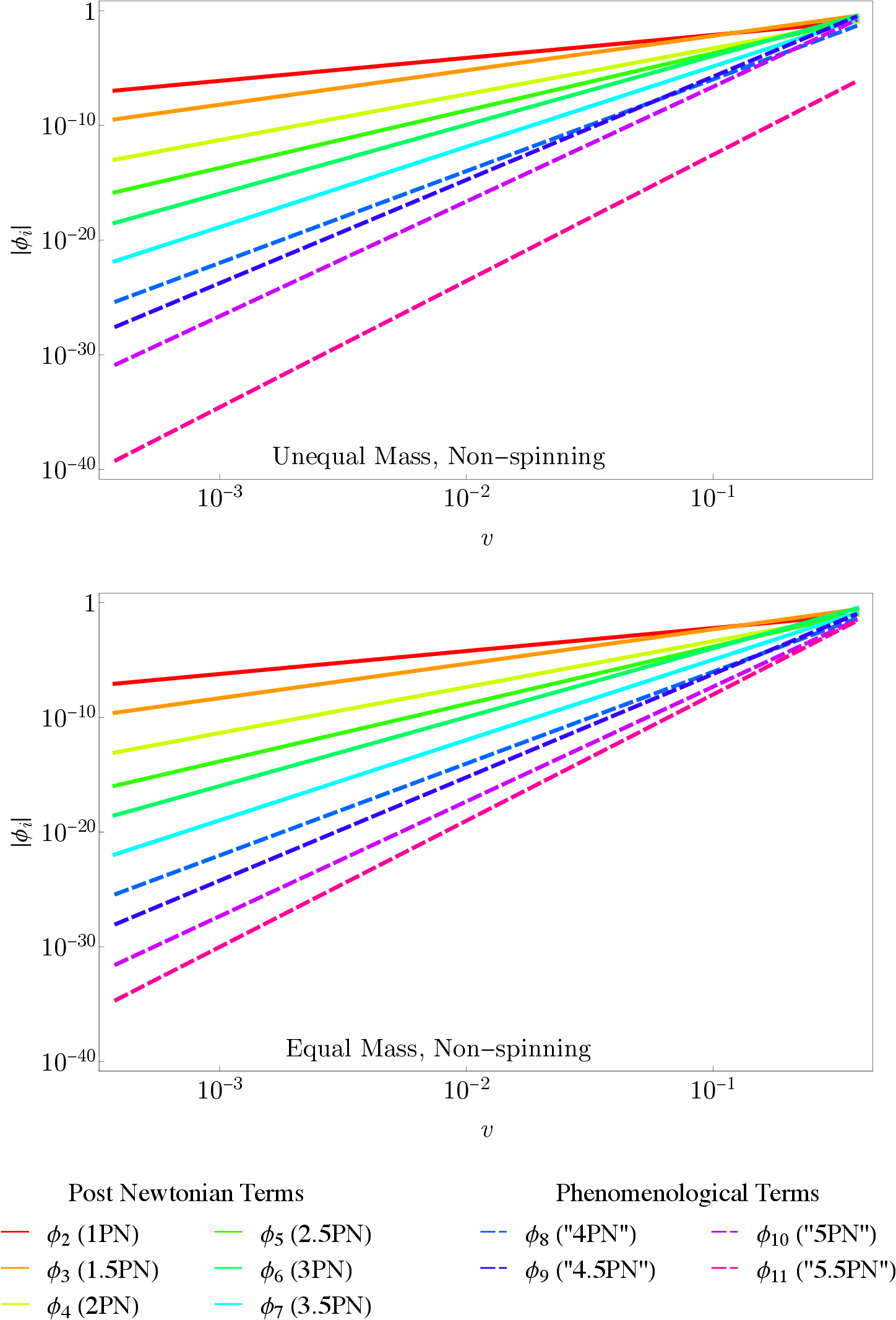} 
  \caption{\label{FIG:allterms} 
  Absolute values of series functions $\phi_i = \phi_i(f,\bm{\theta})$ of the {\tt IMRPhenomD} inspiral phase, as given in Eqs. (\ref{EQ:ins}-\ref{eq:phenomphaseseriesterms}), plotted as a function of velocity $v = (\pi M f)^{1/3}$ for the systems presented in Table \ref{TAB:injections}. Observe that the terms in the phase have a hierarchy, with the lowest PN order terms larger than the higher PN order terms. We do not plot the terms that contain $\log(\pi M f)$ that occur at 2.5PN and 3PN of the classic PN expansion.}
\end{figure}

The {\tt IMRPhenomD} phase during the inspiral 
\begin{align}\label{EQ:ins}
    \phi_\mathrm{Ins}(f,\bm{\theta}) = \phi_\mathrm{TF2}(f,\bm{\theta}) + \phi_{\mathrm{phenom}}(f,\bm{\theta})
\end{align}
is a hybrid model. 
The first term in the above equation comes from the {\tt TaylorF2} model, which derives from the stationary phase approximation of the waveform constructed in a standard Taylor expansion within PN theory \cite{Damour:2002kr}. 
This phase is known to 3.5PN order and it can be expressed as  
\begin{align}\label{EQ:TF2}
\phi_\mathrm{TF2}(f,\bm{\theta}) &=~2\pi f t_c - \varphi_c -\pi/4 
\nonumber\\ 
&~
+(\pi f M)^{-5/3} \sum_{i=0}^7 \phi_i(f,\bm{\theta})\,,
\end{align} 
where the \texttt{TaylorF2} series functions are
\begin{align}
\phi_i(f,\bm{\theta}) = \frac{3}{128\eta}(\pi f M)^{(i/3)}\varphi_i \,,~~~~0\le i\le7,
\end{align}
and the PN coefficients $\varphi_i$ are functions of the masses and the magnitude of the spin angular momenta~\cite{Khan:2015jqa}.
Above,  $M = m_1 + m_2$ is the total mass, and $\eta = m_1 m_2/M^2$ is the symmetric mass ratio, determined by the component masses $m_1$ and $m_2$ where by convention we assume $m_1 \geq m_2$.
The time $t_c$ and phase $\varphi_c$ of coalescence are, for the purpose of data analysis, just a reference time and phase offset.
Since the orbital velocity of the system is $v = (\pi M f)^{1/3}$, we see that $\phi_\mathrm{TF2}(f,\bm{\theta})$ is a Frobenius series in powers of $v$, with a controlling factor that goes as $v^{-5}$.
In a PN series of this type, a term proportional to $v^{2n} \propto f^{2n/3}$ relative to the controlling factor is said to be of $n$PN order. 

The second term in Eq.~\eqref{EQ:ins} contains phenomenological corrections that are expected to enter at 4PN order and above within PN theory.
More specifically, the {\tt IMRPhenomD} model postulates the 5.5PN-accurate phenomenological corrections to the phase 
\footnote{Note that the controlling (leading) factors of the \texttt{TaylorF2} and phenomenological Frobenius series as defined above differ by a factor of $\pi^{-5/3}/128$. In principle, from the basics of PN theory, the controlling factors should be the same, but we will not rescale them here to stay close to historical conventions.
}
\begin{align}\label{EQ:phenom}
\phi_\mathrm{phenom}(f,\bm{\theta}) =
 &(\pi f M)^{-5/3} \sum_{i=8}^{11} \phi_i(f,\bm{\theta})\,,
\end{align}
where the phenomenological series functions are
\begin{align}
\phi_i(f,\bm{\theta})
= \frac{3 \pi^{5/3} }{(i-5)\eta}(M f)^{i/3}\sigma_{i-7} \,,~~~~8\le i\le11,
\label{eq:phenomphaseseriesterms}
\end{align}

The $\sigma_i$ coefficients are unknown functions of the masses and spin angular momenta. 
The {\tt IMRPhenomD} model makes use of an \textit{ansatz} where the $\sigma_i$ are represented as bi-variate polynomials in $\eta$ and $(\chi_{\rm PN}-1)$.
Here, $\chi_{\rm PN}$ is a certain function of the dimensionless spins $\chi_1$ and $\chi_2$, $\eta$ and $M$.
More precisely, the {\tt IMRPhenomD} model represents the $\sigma_i$ coefficients via
\begin{align}
\sigma_i = \sum_{j=0}^{2} \sum_{k=0}^{2} \lambda^i_{jk} \; \eta^j \; \left(\chi_{\rm PN} - 1\right)^k\,.
\label{EQ:PhenomSigma}
\end{align}
The fitting coefficients $\lambda^i_{jk}$ are determined by fitting the waveform phase in the frequency domain to the phase of the Fourier transform of a finite set of numerical relativity simulations and EOB waveforms~\cite{Khan:2015jqa}. 
The fits will be susceptible to statistical fitting error, fitting error due to the finite nature and finite accuracy of the numerical relativity waveforms used to do the fits, and fitting error due to the functional form of the ansatz from  Eq.~\eqref{EQ:PhenomSigma}.
Nonetheless, in the \texttt{IMRPhenomD} model, one picks the best-fit values for these $\lambda^i_{jk}$ coefficients and disregards these fitting uncertainties.
The phenomenological terms of the waveform phase present a similar convergence behaviour to the \texttt{TaylorF2} terms, as shown in Fig.~\ref{FIG:allterms}.
That is, the individual contributions of the 4, 4.5, 5 and 5.5PN terms are smaller than that of proceedings terms, presenting a PN hierarchy up to the upper bound of the inspiral region, which in the \texttt{IMRPhenomD} model is defined at $v_\mathrm{max}/c = (.018 \pi)^{1/3} \sim 0.38$. 
This suggests that we could consider the phenomenological terms as behaving as higher PN order terms, even though they formally do not need to.
In order to make this distinction clear, we will put quotation marks around the PN order of phenomenological terms, for example referring to the $\sigma_4$ term as a ``5.5PN'' term.   
We use this different notation to remind ourselves that the $\sigma_i$ coefficients in the phase are not derived directly from the underlying theory (general relativity), but rather, are obtained from fitting, which in itself carries both systematic and statistical uncertainties. 

%%%%%%%%%%%%%%%%%%%%%%%%%%%%%%%%%%%%%%%%%%%%%%%%%%%%%%%%%%%%%%%%%%%%%%%%%%%%%%%%%%%%%%%%%%%%%%%%%
\section{Injection and Recovery Campaign}\label{SEC:ExperimentalDesign}

In this paper, we set out to understand how PN corrections to the inspiral phase impact systematic error in intrinsic parameters estimated from binary black hole inspirals.
The systematic error present in a given signal will, as detectors improve, eventually become dominant over statistical error, as the latter scales inversely with the SNR of the signal. 
We therefore ask the following question:  As a function of the PN order of the recovery model and the SNR present in an inspiral signal, at what point does the systematic error become larger than the statistical error?
In order to answer this question, we will perform a synthetic injection and recovery campaign. 
In this section, we present the details of the injection and the recovery models and the data analysis methods used. 

Given an advanced LIGO/Virgo observation, one does not know the exact waveform generated by Nature or the exact binary parameters that produced the signal. 
Therefore, the synthetic injected data will be created with the full {\tt IMRPhenomD} inspiral model, for a set of systems detailed in Table \ref{TAB:injections}.
The recovery model will also be the {\tt IMRPhenomD} inspiral model, but truncated at a given PN order (as listed in Table~\ref{TAB:models}), since we have seen that the phenomenological terms represent PN corrections.

Our choice of the {\tt IMRPhenomD} model will not affect the conclusions of this work because both the injection and recovery models are built from this base. 
Therefore, our analysis will be subject to the same inherent model inaccuracies in both the injection and the recovery. 
In turn, this implies that our analysis will only be sensitive to the systematic errors we are introducing ourselves through truncation of the PN terms in the phenomenological Fourier phase. 
A similar analysis could be carried out with the updated {\tt IMRPhenomXAS} model\cite{Pratten:2020fqn}, or extended to include modifications to the merger and ringdown, but we leave this for future work. 

The details of the injection parameters that define the systems we study are the following. 
Our analysis will focus only on the inspiral regime, and therefore, although we consider two separate binary configurations, we choose a total mass of $20 \mathrm{M}_\odot$ for both. 
This guarantees that the SNR is dominated by the inspiral part of coalescence.
One injected configuration will have two bodies of equal mass, while the other will have one body that is three times more massive than the other.
In both cases, we choose to set the spin to zero.
The SNR scales inversely with the luminosity distance $D_L$ of the binary, so for each configuration, we choose injected luminosity distances to obtain the SNRs listed in Table~\ref{TAB:injections}.
We do so to roughly fix the statistical error (which predominantly scales as $1/{\rm{SNR}}$) across injected configurations. 
We select SNRs to encompass the sensitivities of our current (second-generation, 2G) detectors, as well as that of near-future detectors. 
Other details of the system parameters used in the injections can be found in Table \ref{TAB:injections}. 

\begin{table}[th]
\centering
\caption{\label{TAB:injections} Properties of injected binary system configurations. 
Here, $m_i$ is the mass of the $i$-th black hole and $D_L$ is the luminosity distance between Earth and the source that ensures the listed SNR as measured by a current detector network. In addition to the parameters listed in the table below, we set the aligned dimensionless spins $\chi_{1,2}=0$, the inclination angle $\iota = 0.4$ rad, the polarization angle $\psi=2.659$ rad, the right ascension $\mathrm{ra} = 1.375$ rad, the declination $\mathrm{dec}=-1.2108$ rad, the coalescence phase $\phi_c=1.3$ rad, and the geocentric coalescence time $t_c = 1126259642.413$ sec for each injection.}
\begin{tabular}{c c c c c c c}
\hline\hline
& $m_1~(\mathrm{M}_{\odot})$&$m_2~ (\mathrm{M}_{\odot})$&$D_L(\mathrm{Mpc})$& SNR\\
\hline
Equal Mass& 10&10&629.853& 20\\
& 10&10&314.926& 40\\
& 10&10&157.463& 80\\
\hline
Unequal Mass& 15&5&534.780& 20\\
& 15&5&267.390& 40\\
& 15&5&133.695& 80\\
\hline\hline
\end{tabular}
\end{table}

\begin{table}[th]
\centering
\caption{\label{TAB:models} Injection and recovery models used in our analyses. 
The coefficients $\sigma_i$ are the phenomenological coefficients contained in Eq.~\ref{EQ:phenom}. }
\begin{tabular}{c c c}
\hline\hline
&Label & Coefficients set to zero\\
\hline
{\raggedleft Injection model:}&``5.5PN'' & none \\
\hline
{\raggedleft Recovery models:}&``5.5PN'' & none \\
&``5PN'' &$\sigma_4=0$ \\
&``4.5PN''&$\sigma_3=\sigma_4=0$\\
&``4PN''&$\sigma_2=\sigma_3=\sigma_4=0$\\
&3.5PN&$\sigma_1=\sigma_2=\sigma_3=\sigma_4=0$\\
\hline\hline
\end{tabular}
\end{table}

The SNR and the results of our parameter estimation studies will depend on the spectral noise density of the detectors assumed to have measured the synthetic injected signals. 
To represent a current detector network, we focus on a set of ground-based detectors, comprised of LIGO Hanford \cite{LIGOScientific:2014pky}, LIGO Livingston \cite{LIGOScientific:2014pky}, and Virgo \cite{TheVirgo:2014hva}. 
In particular, we use analytic approximations for the design sensitivities of advanced LIGO and advanced Virgo \cite{OReilly_2022}, which should approximately correspond to that of the fourth observing run \cite{KAGRA:2013rdx}. 

To perform the parameter estimation analyses, we use the Bayesian inference library BILBY \cite{Ashton:2018jfp, Romero-Shaw:2020owr}. 
This library obtains its waveform approximants from the LALSimulation \cite{lalsuite} package and provides access to several Bayesian inference packages, including {\tt dynesty} \cite{Speagle_2020}, the nesting sampling package that we use for this project. 
In each parameter estimation study, we vary over all model parameters of the \texttt{IMRPhenomD} model except for the dimensionless spins, namely chirp mass, $\mathcal{M}$, mass ratio $q$, luminosity distance $D_L$, right ascension $\mathrm{ra}$, declination $\mathrm{dec}$, inclination angle $\iota$, polarization angle $\psi$, coalescence phase $\phi_c$, and coalescence time $t_c$. 
We choose not to vary over the dimensionless spins because we are considering only non-spinning systems and the reduction in parameter space allows for faster analyses. While there is degeneracy between the mass and spin parameters, this simplification should not change the qualitative conclusions of this study. 
All analyses are performed using 1024 live points for each of the three duplicate analyses done for each model variant.
Otherwise, our analyses settings (both sampling configuration, and prior choices) match those from recent LIGO/Virgo/KAGRA publications~\cite{LIGOScientific:2021usb, LIGOScientific:2021djp, ligo_scientific_collaboration_and_virgo_2022_6513631, ligo_scientific_collaboration_and_virgo_2021_5546663}.
After exploring the $9$-dimensional likelihood surface, we will present a subset of corner plots (to avoid crowding the presentation), focusing only on some astrophysically relevant parameters. 

With these details in hand, we perform parameter estimation analyses on each injection configuration with each recovery model, leaving us with 30 analyses to complete. 
When injecting the data and exploring the likelihood surface, we set an upper bound on the frequency at the end of the inspiral region ($Mf_\mathrm{max}=0.018$) to isolate the impact of modifying the inspiral model. 
For the binaries considered here with a total mass of $M = 20 \mathrm{M}_\odot$, this corresponds to a frequency of $f_\mathrm{max} = 182.7 \mathrm{Hz}$. 

The truncation of the frequency range used is applied only to the evaluation of the likelihood itself without any additional termination conditions applied to the underlying waveforms used, this in order to avoid parameter biases introduced by such unphysical sharp waveform features as explored in~\cite{Rodriguez:2013mla, Mandel:2014tca}.
We will comment later on how our results are impacted by a different choice of maximum frequency. 

%%%%%%%%%%%%%%%%%%%%%%%%%%%%%%%%%%%%%%%%%%%%%%%%%%%%%%%%%%%%%%%%%%%%%%%%%%%%%%%%%%%%%%%%%%
\section{\label{SEC:Results}Systematic and Statistical Uncertainties through Bayesian Parameter Estimation}

In this section, we present the results of the Bayesian parameter estimation analyses laid out in Sec.~\ref{SEC:ExperimentalDesign}. In particular, we will focus on the systematic errors introduced into the chirp mass $\mathcal{M} = \eta^{3/5} M$ and the mass ratio $q = {m_2}/{m_1} < 1$  due to truncation of the inspiral {\texttt{IMRPhenomD}} model. As we will see, strong correlations with the reference time $t_c$ will also force us to include this parameter in the partial corner plots we present to explain our results. We emphasize again that even though we present corner plots for only a subset of the parameters of the \texttt{IMRPhenomD} model, we do vary over all the parameters of the model (except the spins), as discussed in the previous section.  

\begin{figure*}[ht]
\includegraphics[width=0.485\textwidth]{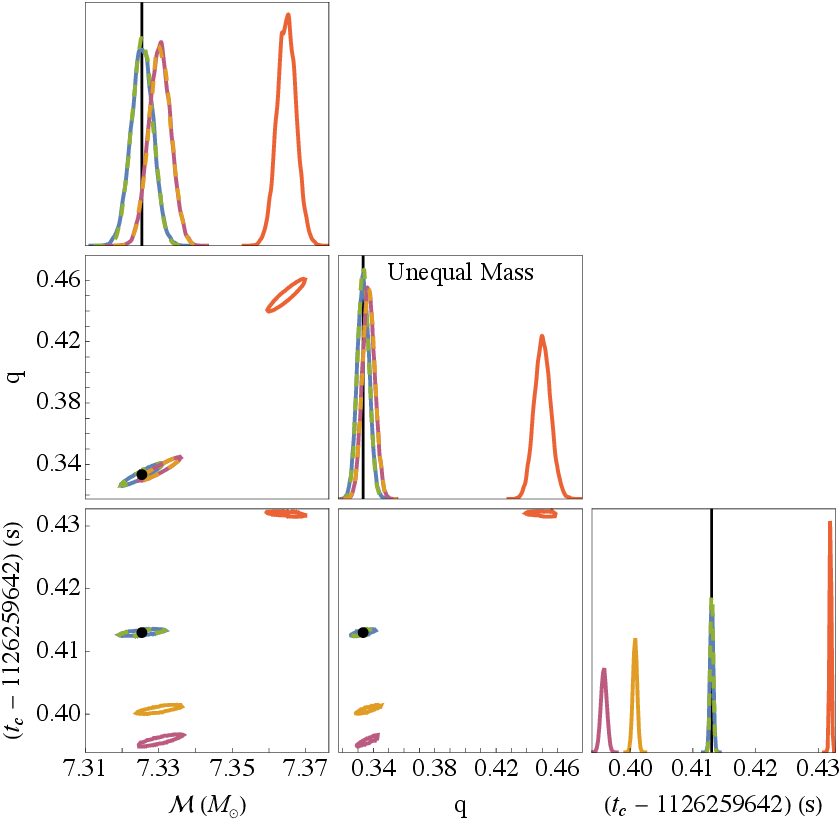} \quad
\includegraphics[width=0.485\textwidth]{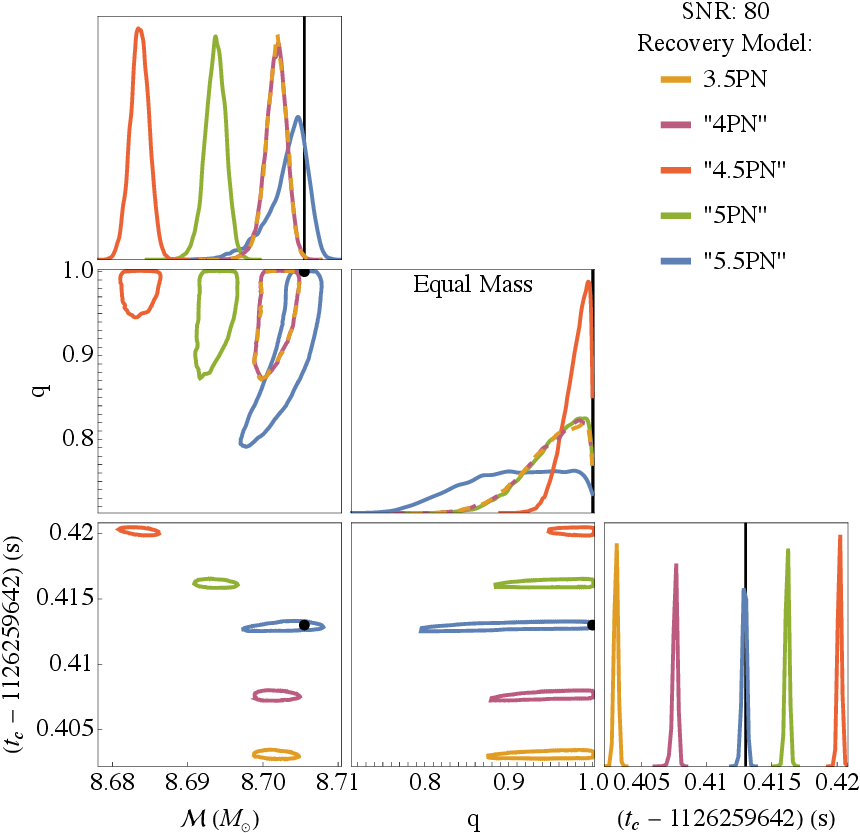}
\caption{\label{FIG:q3_model_corner}
Partial corner plot produced after a Bayesian parameter estimation study, using each of the recovery models detailed in Table~\ref{TAB:models} to infer parameters from the unequal mass (left) and equal mass (right) SNR 80 injected signals, described in Table \ref{TAB:injections}. 
The corner plots show the marginalized one-dimensional posterior probability distributions and the 90\% credible region contours of the two-dimensional posterior probability distributions on chirp mass $\mathcal{M}$, mass ratio $q$, and (geocentric) reference time $t_c$.
The injected ``true'' values are shown as black dots and vertical lines.
Observe that the bias is very small for the ``5.5PN'' model in both cases because the recovery model is the same as the injection model.  
The bias is also suppressed for the ``5PN'' model in the unequal mass case where the removed ``5.5PN'' term is very small.
However, in the equal mass case where the ``5.5PN'' term has a larger contribution, the bias from the ``5PN'' model is significant.
The bias of the ``4.5PN'' model is very large in both cases, because of the alternating structure of the phenomenological coefficients. 
The bias becomes smaller again, but is still not negligible, for the 3.5PN and ``4PN'' models. 
This suggests that the phenomenological coefficients of the \texttt{IMRPhenomD} model are important for parameter estimation and their inaccurate determination could lead to systematic bias. 
}
\end{figure*}

Let us first consider the unequal mass injected signal with SNR = 80.
The left panel of Fig.~\ref{FIG:q3_model_corner} presents the partial corner plot obtained from our Bayesian parameter estimation analyses, using each of the recovery models detailed in Table \ref{TAB:models}. The figure depicts one-dimensional marginalized posterior probability distributions and the 90\% credible region contours of the marginalized two-dimensional posterior distribution.  
While our focus is on $\mathcal{M}$ and $q$, we also include the reference (geocentric) coalescence time $t_c$, because its correlation with mass parameters helps to explain the results.  
Observe that the posterior probability distributions obtained using the ``5.5PN'' model are centered on the injected values, indicated with black vertical lines. 
This is as expected because the recovery model is identical to the one used to produce the signal injection. 
Therefore, there is no systematic error introduced by the model and any small bias present is due to sampling error. 
Observe also that the ``5PN'' model performs similarly well, this behavior can be explained from Fig. \ref{FIG:allterms}.
For the unequal mass injection, the magnitude of the ``5.5PN'' term is significantly smaller than that of any other term.  
Therefore, the systematic error introduced by removing this term from the model is negligible and again the small bias is dominated by sampling error. 

The 3.5PN, ``4PN'' and ``4.5PN'' models, however, show different behavior in the left panel of Fig.~\ref{FIG:q3_model_corner}. Both the 3.5PN and ``4PN'' models produce a slight bias from the injected values in the mass parameters, with very similar posterior probability distributions. On the other hand, the posterior probability distributions obtained with the ``4.5PN'' model indicate a large bias with negligible support at the injected values. 
At first glance, these results are surprising, because one would na\"ively expect the ``4.5PN'' model to be more accurate, and thus produce a smaller bias, than the 3.5PN and ``4PN'' models. 
However, some of the phenomenological coefficients have alternating signs: while the ``4PN'', ``5PN'' and ``5.5PN'' terms are all positive on the relevant range of parameters, the``4.5PN'' term is negative (Fig~\ref{FIG:allterms} plots absolute values) but of similar magnitude to the ``4PN'' term. 
This means that the ``4.5PN'' term approximately cancels the ``4PN'' term,  and so truncating the model at this order produces a less accurate model. 
Once this term is removed to produce the ``4PN'' model we are able to more accurately recover the mass parameters. 

This alternating behavior of the series is not just a feature of the phenomenological coefficients but also a known feature of the classic PN expansion. Indeed, it has been known for a while that parameter estimation with a 2.5PN model is less accurate than with a 2PN model, even though the former is formally more accurate \cite{Poisson:1995vs}. This is precisely because of the alternating structure of the series. In the case of the \texttt{IMRPhenomD} model, however, the higher than 3.5PN order coefficients are all phenomenological, so it is not clear that this alternating sequence should continue, or that the magnitude of the terms in all of parameter space have been accurately determined through fitting. This can only be determined by either calculating the higher PN order term, or re-doing the fits with a denser set of numerical relativity simulations and quantifying the statistical uncertainty of the fits. 

The similarity between the one-dimensional posterior probability distributions in the left panel of Fig.~\ref{FIG:q3_model_corner} found using the 3.5PN and ``4PN'' models can be explained by studying the Fourier phase model of Eqs.~\eqref{EQ:ins}-\eqref{EQ:phenom}.
The total phase includes the term $2\pi f t_c$, which has the same linear dependence on frequency as the ``4PN'' term when accounting for the control factor of the PN expansion. 
The presence of this term compensates for the removal of the ``4PN'' term and allows the 3.5PN model to recover similar mass parameters at the cost of biasing $t_c$. 
We see this in the left panel of Fig.~\ref{FIG:q3_model_corner}: while the posteriors in the mass parameters coincide, the posteriors on $t_c$ show a large bias between the two models.

The same qualitative conclusions hold for other mass ratios, although for equal-mass injections one can encounter boundary effects from the prior, as shown in the right panel of Fig.~\ref{FIG:q3_model_corner}. The right panel is analogous to the left panel of Fig.~\ref{FIG:q3_model_corner}, but for the equal-mass binary with the same SNR. 
The posterior probability distributions for $\mathcal{M}$, for example, are very similar to that of the unequal mass case.
However, because there is not such a large difference between the ``5.5PN'' and ``5PN'' terms, as shown in Fig. \ref{FIG:allterms}, we see that the systematic bias in $\mathcal{M}$ grows with the ``5PN'' and ``4.5PN'' models, and then decreases for the 3.5PN and ``4PN'' models.

The right panel of Fig.~\ref{FIG:q3_model_corner} also shows that the posterior probability distribution for $q$ is affected by the boundary of the $q$ prior (at $q=1$), and thus it provides less information about systematic bias.
This boundary is largely artificial, as it is set by the convention of $m_1 \geq m_2$, but nonetheless, it reduces the accessible parameter space. 
While, for the unequal-mass binary, a mismatch between the full and truncated \texttt{IMRPhenomD} waveforms could be ``corrected'' by introducing a bias that either increases or decreases the mass ratio around the true value, the equal-mass case is only provided with freedom in one direction.
\begin{figure}[ht]
\includegraphics[width=0.45\textwidth]{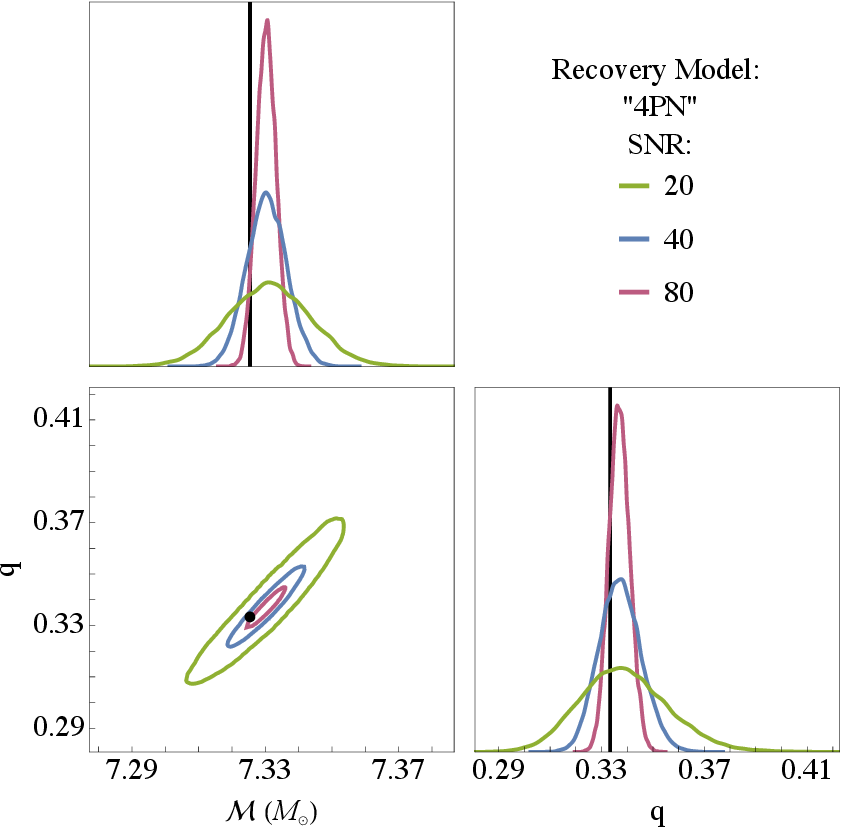}
\caption{\label{FIG:snr_corner} Partial corner plot produced using the ``4PN''  recovery model detailed in Table~\ref{TAB:models} to estimate parameters from each of the unequal mass injected signals described in Table~\ref{TAB:injections}.
We depict the marginalized one-dimensional posterior probability distributions and  90\% credible region contours of the two-dimensional posterior probability distributions on chirp mass $\mathcal{M}$ and mass ratio $q$.
The injection values are indicated in black.
Observe that, as expected, as the SNR increases, the width of the posterior probability distribution shrinks, reducing the support of the posterior on the injected values. 
This figure indicates clearly that models once thought to be accurate enough at small SNR can rapidly become insufficiently accurate at higher SNRs.}
\end{figure}

Let us now study how the above conclusions are affected by the SNR of the signal. Figure \ref{FIG:snr_corner} illustrates the impact of the SNR on the relationship between systematic and statistical error, focusing on only three recoveries, all with the ``4PN'' model but each with a signal of different SNR.
As expected, at the lower SNRs, the injected values are contained within the credible contours of the recovered posterior probability distribution. 
However, at the highest SNR, this is no longer the case and the posterior probability distributions have less support at the injected values. 
Clearly then, as the SNR of the observed signals increases, models that were once accurate enough to recover parameters reliably are no longer sufficient, as anticipated.    

\begin{figure*}[t]
\includegraphics[width=0.45\textwidth]{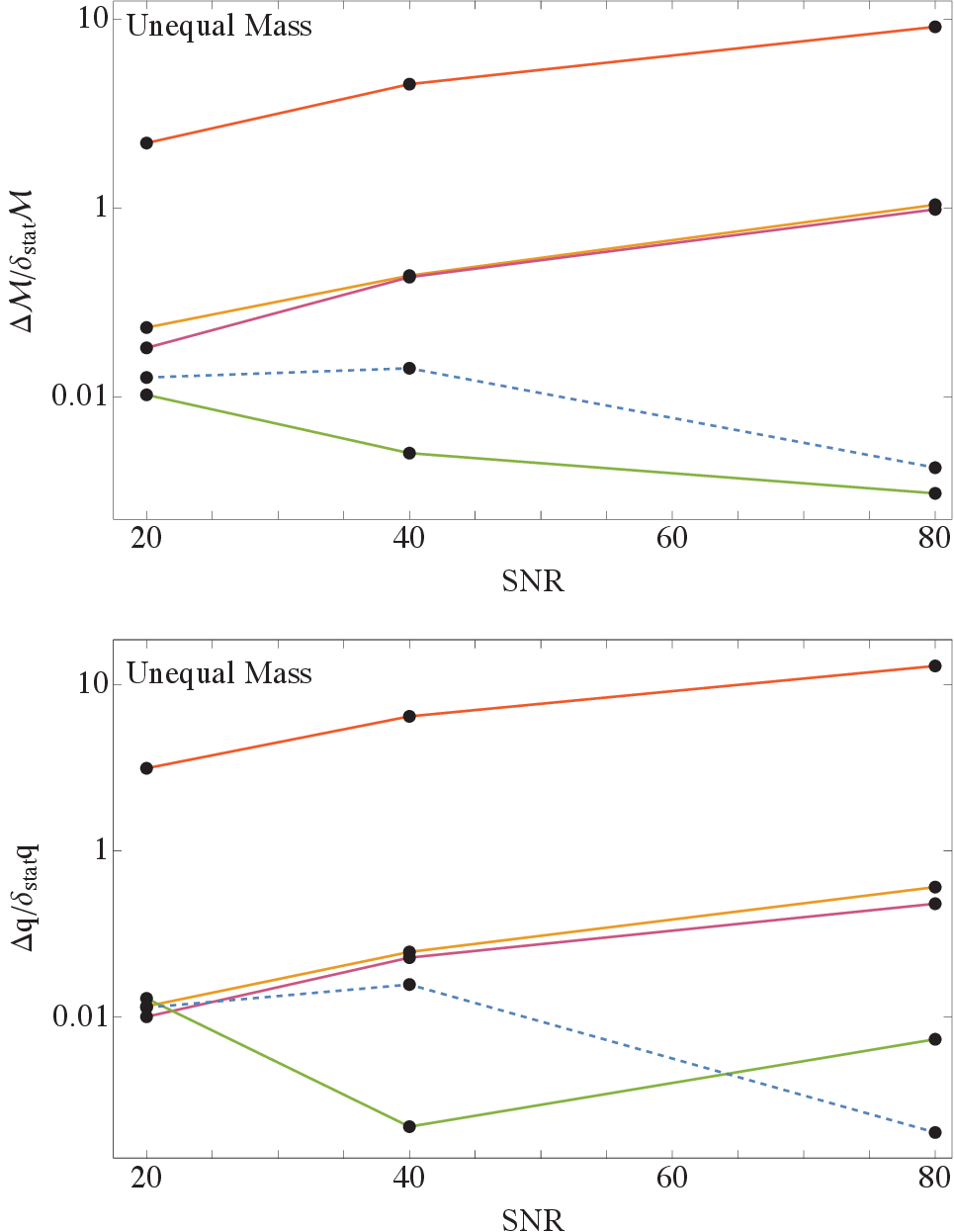} 
\hspace{0.5cm}
\includegraphics[width=0.45\textwidth]{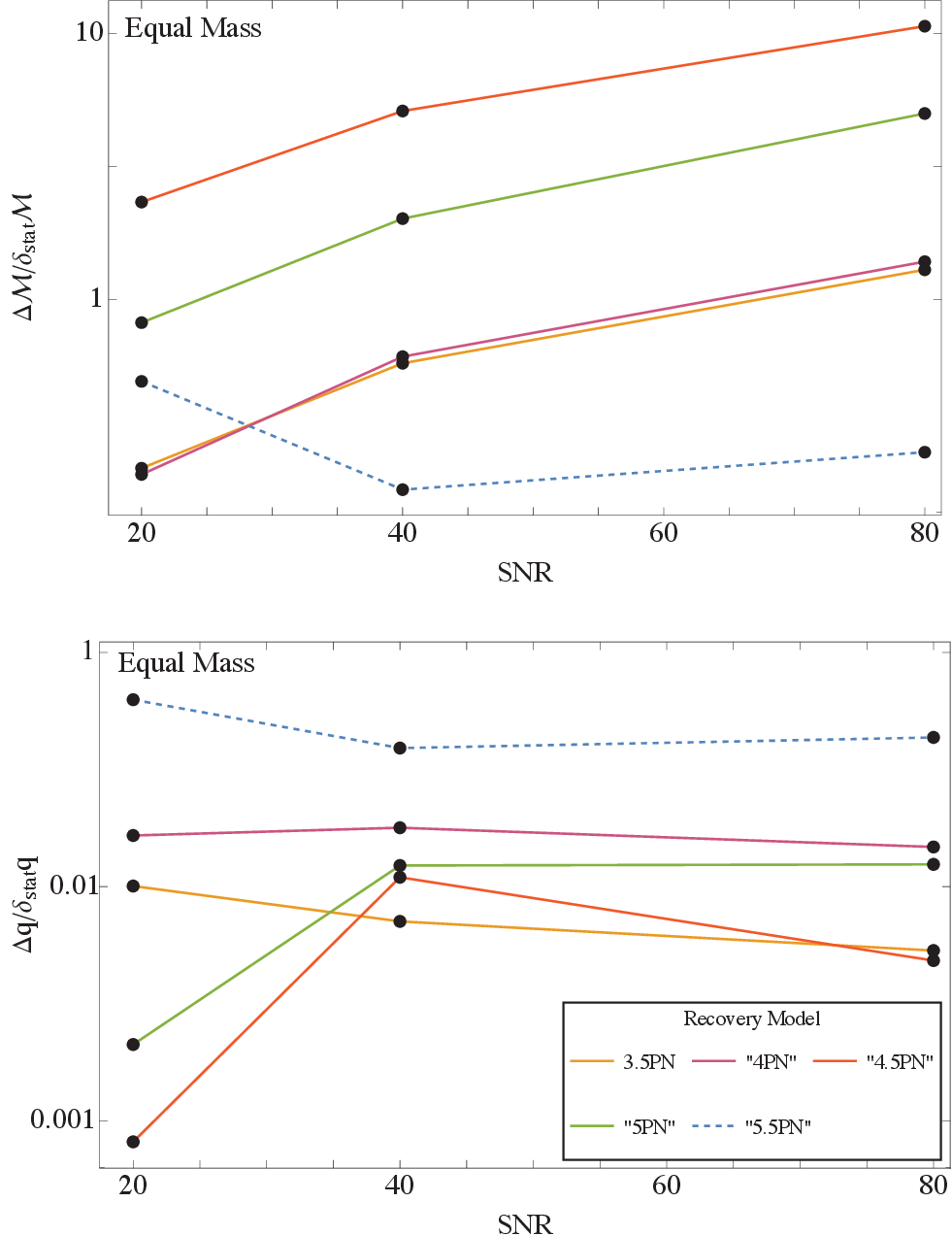}
\caption{\label{fig:q3results} Ratio of bias to statistical error in the chirp mass (top) and mass ratio (bottom) for various recovery models (Table~\ref{TAB:models}) and injected parameter (Table~\ref{TAB:injections}), including the unequal mass cases (left) and the equal mass cases (right). 
The black dots indicate the configurations of the parameter estimation analyses we performed, which we join through a nearest-neighbor linear interpolation. 
When the recovery model matches the injection model (i.e.~in the ``5.5PN'' case), the bias is due entirely to sampling, and it is thus distinguished by plotting it with dashed lines.
Observe that for the 3.5PN, ``4PN,'' and ``4.5PN'' recovery models in the unequal mass case and 3.5PN, ``4PN,''  ``4.5PN,''  and ``5PN'' recovery models in the equal mass case, where the bias is dominated by systematic error, the bias grows with SNR relative to the statistical error. 
Observe that this trend is violated for the mass ratio in the equal-mass case due to prior boundary effects. 
Observe also that systematic bias grows as ``PN'' terms are removed, until the ``4.5PN'' term is removed.
Most significantly, note the instances where $\Delta\theta/\delta_\mathrm{stat} > 1 $.
In these cases, the recovery model was not accurate enough to reliably recover the injection parameters.
}
\end{figure*}

Let us now collect all of our results and study how the systematic and statistical uncertainties behave with SNR for different recovery models, as shown in Fig.~\ref{fig:q3results}.
To do so, we take the statistical error $\delta_\mathrm{stat}\theta$ on a given parameter $\theta$ to be half of the 90\% credible interval on the one-dimensional marginalized posterior probability distribution on that parameter.
We then take the bias in the recovered parameter to be $\Delta\theta = |\theta_\mathrm{rec} - \theta_\mathrm{inj}|$, where $\theta_\mathrm{inj}$ is the injected value and the recovered value $\theta_\mathrm{rec}$ is determined by the maximum of the full nine-dimensional posterior probability distribution
\footnote{It is essential that the set of recovered parameters $\bm{\theta}_\mathrm{rec}$ is taken from the maximum of the full $N$-dimensional posterior probability distribution (where $N$ is the number of parameters varied in the Bayesian analysis) rather than the maximum of marginalized one- or two-dimensional posterior probability distributions such as those depicted in Fig \ref{FIG:q3_model_corner}. While we have verified that the peaks depicted in the corner plots in this paper correspond closely to the peaks of the full posterior probability distribution, this need not always be the case. Especially in the case of highly correlated parameters, it is possible that the maximum of a marginalized posterior probability distribution does not line up with the maximum of the full posterior probability distribution.}. 
Figure~\ref{fig:q3results} shows $\Delta\theta/\delta_\mathrm{stat}$ as a function of SNR of the injection, for each recovery model and for both the unequal mass (left) and equal mass (right) systems, with $\theta = \mathcal{M}$ (top) and $\theta = q$ (bottom). 
In the case of the equal mass configuration in the right panels of Fig. \ref{fig:q3results},
we use a one-sided credible interval as the statistical error 
because the injected value $q=1$ is at the boundary of the allowed range.

This figure can be read and interpreted as follows. 
The bias for a given parameter and recovery model is identical to the statistical error when $\Delta\theta/\delta_\mathrm{stat}\theta=1$.
When $\Delta\theta/\delta_\mathrm{stat}\theta1<1$, the injected value of the parameter falls within the confidence intervals of the recovered posterior probability distribution, which one interprets as a sign of trust in the inferred point-estimate value. 
However, when $\Delta\theta/\delta_\mathrm{stat}>1$, the injected parameter is outside of the recovered confidence intervals.
When this is the case, one interprets that the recovery models is not accurate enough to reliably recover the injection parameters. 

Figure~\ref{fig:q3results} allows us to make several observations. 
When the recovery model matches the injection model, as is the case for the ``5.5PN'' recovery model, there is no mis-modeling error and $\Delta\theta$ is entirely produced due to sampling error. 
We include the recoveries done with this model in our results to demonstrate the bias that can be expected even in the absence of systematic error (but we represent it with dashed lines to emphasize its difference from the recoveries done with the other models). 
As shown in the figure, the sampling error of the ``5PN'' model is comparable to the systematic error of the ``5.5PN'' model in the unequal mass case, because, as stated before, the ``5.5PN'' term in this case is very small. 
In the equal-mass case, however, the systematic error of the ``5PN'' model is much larger than the sampling error, and grows with SNR. 
Indeed, the ratio of the systematic error to the statistical error tends to grow with SNR for all recovery models, except for those referring to the mass ratio in the equal mass case (due to prior boundary effects). 
Another trend we observe is that, as expected from Fig.~\ref{FIG:q3_model_corner}, systematic bias grows as ``PN'' terms are removed, until the ``4.5PN''model, which leads to the most systematic error for the reasons explained earlier. 
Additionally, because of correlations with the $t_c$ term, which as explained before enters at an effective 4PN order, the 3.5PN and ``4PN'' recovery models produce similar biases. 
 
To verify that the biases seen here are in fact due to the truncation of the higher PN corrections in our recovery models, we performed a set of analyses where we lowered the maximum frequency of our injected signal from $M f_{\rm max} = 0.018$ to $M f_{\rm max}  = 0.01$.
We focused on our SNR=80 unequal mass injection and performed two analyses with a lowered maximum frequency: one where all injection parameters are kept the same, resulting in an SNR $<$ 80 signal of shorter duration, and one where the luminosity distance $D_L$ is rescaled to maintain an SNR of 80 while leaving all other parameters unchanged.
We found that, in lowering the maximum frequency, the contribution to the signal from the higher order terms is reduced, as expected. 
Moreover, the biases in inferred parameters become significantly smaller for the shorter signal, indicating that the biases in our original analyses are a result of the truncated PN approximation.
However, we also find that the width of posterior probability distributions increases for the shorter signals, consequently increasing the statistical error.
This increase is seen in both the analyses where $D_L$ is left unchanged and where it is rescaled to maintain an SNR of 80. 
The reason for this is that statistical error scales inversely with SNR only approximately; in reality, the likelihood surface can have multiple valleys and peaks, none of which need to be perfectly Gaussian (as assumed when deriving the $1/{\rm{SNR}}$ scaling of the statistical error with a Fisher analysis). 
The structure of the likelihood, in turn, depends on the duration of the signal, since shorter signals will allow for stronger correlations and more structure.

The biases introduced during the inspiral by truncating PN corrections to the phase do, of course, depend on the total mass  $M$ of the injected signal.
If the power spectral density were flat, then the biases would be total mass insensitive because the inspiral frequency cutoff  $f_{\rm max} \propto 1/M$ and so $v_{\rm max}  = (0.018\pi)^{(1/3)}$ is independent of total mass. 
The power spectral density, however, is very much not flat, and it rises at lower frequencies. 
This implies that higher total mass signals spend less of their inspiral in band, until eventually, at a sufficiently high total mass, only the merger and ringdown are observed. 
Shorter in-band inspirals will lead to a less accurate posterior recovery and larger correlations with other parameters, as the signal will contain less information. 
This is why in our analysis we selected a total mass of $M=20 \mathrm{M}_\odot$ for all injections, so as to ensure that the SNR of the observed signal would be dominated by the inspiral, allowing us to disregard the merger and ringdown.

A useful way to quantify the disagreement between two waveforms $h_1$ and  $h_2$ is the mismatch
\begin{align}\label{EQ:MM}
    MM(h_1,h_2) = 1 -  \max\limits_{t_c,\phi_c}  \left[\frac{(h_1|h_2)}{\sqrt{(h_1|h_1) (h_2|h_2)}}\right],
\end{align}
where the inner product is defined as
\begin{align}
    (h_1|h_2) = 4 \mathrm{Re}\int_{0}^{\infty} \frac{\tilde{h_1^*}\tilde{h_2}}{S_n(f)}df,
\end{align}
with $S_n(f)$ the noise power spectral density of the detector and the asterisks denoting complex conjugation. The mismatch takes values between zero and unity, with $MM=0$ corresponding to perfect agreement between $h_1$ and $h_2$ up to time and phase offsets (see \cite{Moore:2018kvz} for a useful discussion on computing mismatch for quasi-circular and eccentric waveform models). Even with perfect waveform models the expected value of the mismatch between injected and recovered waveforms is non-zero due to noise. For signals in stationary, Gaussian noise, the expected mismatch is ${\rm E}[MM] = (D-1)/(2\, {\rm SNR}^2)$, where $D$ is the number of parameters that describe the waveform model~\cite{Chatziioannou:2017tdw}. Note that this statistical error in the match scales inversely with the square of the signal-to-noise ratio, so as detectors become more sensitive, the statistical error drops very rapidly, potentially exposing various sources of systematic error from waveform mis-modeling.
Bias in inferred parameters is produced as a result of disagreement between the injection and recovery models. We can therefore use mismatch as a tool to predict the relative magnitude of the bias a given recovery model will cause and in doing so analytically corroborate the Bayesian results presented in this section. 

We compute the mismatch between our injected signal $h_{\text{``5.5PN''}}(\bm{\theta}_\mathrm{inj})$ and our recovery models injected with with our injection parameters $h_{\mathrm{x}}(\bm{\theta}_\mathrm{inj})$, $\mathrm{x} = $ \{3.5PN, ``4PN'',``4.5PN'',``5PN''\}. When $\mathrm{x} = $ ``5.5PN'', the mismatch is identically zero. We expect that the relative sizes of these mismatches should correlate to the relative sizes of the biases in recovered parameters produced by these models. 

\begin{figure}[ht]
\includegraphics[width=0.475\textwidth]{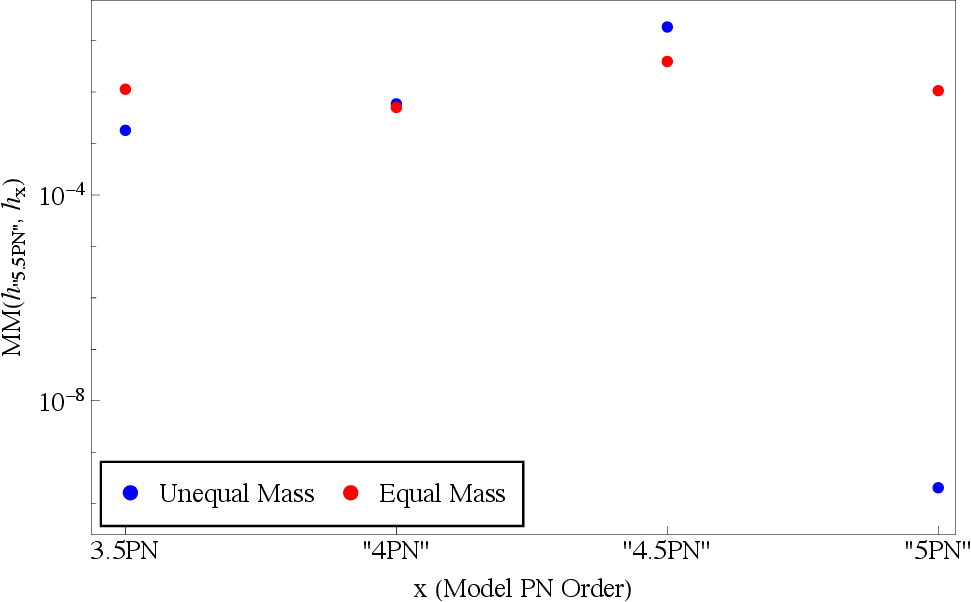}
\caption{\label{fig:missmatch} Mismatch, as defined in Eq. \ref{EQ:MM}, between the injected signal $h_{\text{``5.5PN''}} = h_{\text{``5.5PN''}}(\bm{\theta}_\mathrm{inj})$ and signals produced by injecting the recovery models with the injection parameters $h_{\mathrm{x}} = h_{\mathrm{x}}(\bm{\theta}_\mathrm{inj})$.
We consider the SNR = 80 unequal and equal mass injection detailed in Table \ref{TAB:injections}. Observe that the relative magnitudes of the mismatch values are comparable to the relative magnitudes of the biases in chirp mass and mass ratio depicted in the corner plots in Fig. \ref{FIG:q3_model_corner}} 
\end{figure}

We consider $\bm{\theta}_\mathrm{inj}$ as given by the unequal and equal mass injections at SNR = 80 detailed Tab. \ref{TAB:injections}. In Fig. \ref{fig:missmatch}, we plot $MM[h_{\text{``5.5PN''}}(\bm{\theta}_\mathrm{inj}),h_{\mathrm{x}}(\bm{\theta}_\mathrm{inj})]$. Indeed, we see that the relative magnitudes of the mismatches closely correspond to the relative magnitudes of the biases in mass ratio and chirp mass depicted in Fig \ref{FIG:q3_model_corner}. Notably, we see that the ``5PN'' recovery model in the unequal mass case most closely matches the injected signal, while the largest mismatch in both cases is produced by the ``4.5PN'' model. Biases in the reference time are not captured by the mismatch because it is maximized over this parameter. Through this analysis, we illustrate the connection between bias in the recovered parameters and mismatch between the injection and recovery models and we lend additional credence to the conclusions drawn in this paper. 

%%%%%%%%%%%%%%%%%%%%%%%%%%%%%%%%%%%%%%%%%%%%%%%%%%%%%%%%%%%%%%%%%%%%%%%%%%%%%%%%%%%%%%%%%%%%%%%%%
\section{Marginalizing over uncertainties}\label{SEC:MissingTerms}

The most striking aspect of the results presented in the previous section is the magnitude of the systematic error that we found. 
As we discussed in that section, we considered relatively low total mass systems (for which the total SNR from the whole coalescence is dominated by the inspiral region) and we truncated higher PN-like corrections that we have shown to be appropriately small.
Despite this, the posterior distributions obtained indicate significant biases in the recovered mass parameters, often with very little support at the injected values.
These biases in chirp mass and mass ratio will translate to similarly large biases in the inferred component masses. 
Most notably, these biases occur even at SNRs expected in the next observing runs with current detectors, and will certainly become more significant at higher SNRs with for third-generation detectors.
Of course, systematic bias is not necessarily going to be this large with \texttt{IMRPhenomD}-type models where one never sets the terms in the phenomenological phase to zero. 
However, the phase is indeed truncated at ``5.5PN'' order and the phenomenological coefficients do contain systematic fitting errors (due to the reasons discussed in Sec. \ref{SEC:The Model}). 
In this section, therefore, we search for, propose and develop a method to ameliorate systematic uncertainties due to unknown higher PN order terms.  

Previous attempts at mitigating waveform inaccuracies have shown great promise, but are often limited in scope and applicability~\cite{Moore:2015sza, Doctor:2017csx, Edelman:2020aqj, Jan:2020bdz, Hu:2022rjq}.
Such attempts either only account for and correct model uncertainties in a subset of the full parameter space covered by the full analyses, or provide diagnostics without suggested improvements for the full parameter space. 
A method that is capable of both full coverage and general directive corrections is yet to emerge, but an important first step is presented in~\cite{Read:2023hkv} basing the model for the waveform corrections around the noise-dependent indistinguishability between the waveform models under comparison.

Our proposed method will differ from these previous attempts because our philosophy will be to ``parameterize our ignorance and then marginalize over it.'' In essence, what we propose is that, for any waveform that contains systematic inaccuracies, one should do the following: 
\begin{enumerate}
    \item [(i)] introduce a model for these inaccuracies, characterized by a set of parameters $\bm{\lambda}$, 
    \item [(ii)] explore the likelihood by varying over all parameters ($\bm{\theta} \cup \bm{\lambda}$), 
    \item [(iii)] marginalize over $\bm{\lambda}$ to produce reduced corner plots that range over $\bm{\theta}$ only. 
\end{enumerate}
The expected end result will be to ameliorate the systematic uncertainty at the cost of broadening the posterior probability distributions and thus increasing the statistical error. 

A key element in this method is to properly model our ignorance. For the case considered in this study, our ignorance is entirely encapsulated by unknown higher PN order terms. For the \texttt{IMRPhenomD} model, this means inaccuracies (a) in the ``4PN'' to ``5.5PN'' order terms of the phenomenological phase (due to fitting inaccuracies), and (b) inaccuracies in the higher than ``5.5PN'' order terms (due to truncation). One way to address (a) is to promote the $\lambda^i_{jk}$ constants as new parameters of the \texttt{IMRPhenomD} model, with priors taken to be the posterior probability distributions of the $\lambda^i_{jk}$ fits (as opposed to simply picking numbers for these $\lambda^i_{jk}$, which would correspond to delta-function priors). This method is similar in spirit to how the equation-of-state sensitivity of approximately universal relations~\cite{Yagi:2013bca,Yagi:2013awa} is taken into account when extracting the mass and radius of neutron stars from binary inspirals~\cite{Chatziioannou:2021tdi,LIGOScientific:2018cki}. 

Let us now focus on how to address inaccuracies due to truncation (case (b) above). The best way to model this ignorance is to use our analytic knowledge of the terms that are being ignored. For example, if we have a model that is accurate to 5.5PN order, then the next term in the series will scale as 
$[1/\eta (M f)^{-5/3}] \bar{\sigma}_5 (M f)^{4}$, where the term in square brackets is the controlling factor of the phenomenological approximation. Technically, the 6PN term could also scale as 
$[1/\eta (M f)^{-5/3}] \bar{\sigma}_5 (M f)^{4} \log{M f}$ (due to certain tail terms that arise in the PN approximation); however, the log-correction is mild and similar enough to the non-log term that, for the purpose of marginalizing over our ignorance, either model will suffice. One could, of course, include not just the 6PN term, but also the 6.5PN term simultaneously, and maybe also higher PN order terms. How many terms one must keep will depend on the accuracy of the base model and the SNR of the signal. 

An important question now presents itself: how does one set the priors on these new nuisance parameters $\bar{\sigma}_i$? One option would be to use completely uninformative (i.e.~flat) priors with infinite (or very large) boundaries, but this is not smart. If one were to do this, then some region of the prior would allow sufficiently large values of $\bar{\sigma}_i$, which would render the ignorance term $[1/\eta (M f)^{-5/3}] \bar{\sigma}_5 (M f)^{4}$ much larger than terms at lower (and known) PN order. In fact, for sufficiently large $\bar{\sigma}_i$, the ignorance term would dominate over the rest of the known PN series, rendering the entire approximation invalid. A better choice of prior is to use our analytic knowledge of the structure of our ignorance. Even though we do not know the $\bar{\sigma}_i$ precisely, we do know that they are functions of the system parameters $\bm{\theta}$ and we know the values of $\sigma_i$ at lower (known) PN orders. We can then infer the range that $\bar{\sigma}_i$ can have based on these lower PN terms, and use this as the boundary of a flat prior. Such a methodology was recently used successfully to study tests of general relativity with the LVK implementation of the parameterized post-Einsteinian model~\cite{Perkins:2022fhr}. 

\begin{figure}[ht]
\includegraphics[width=0.475\textwidth]{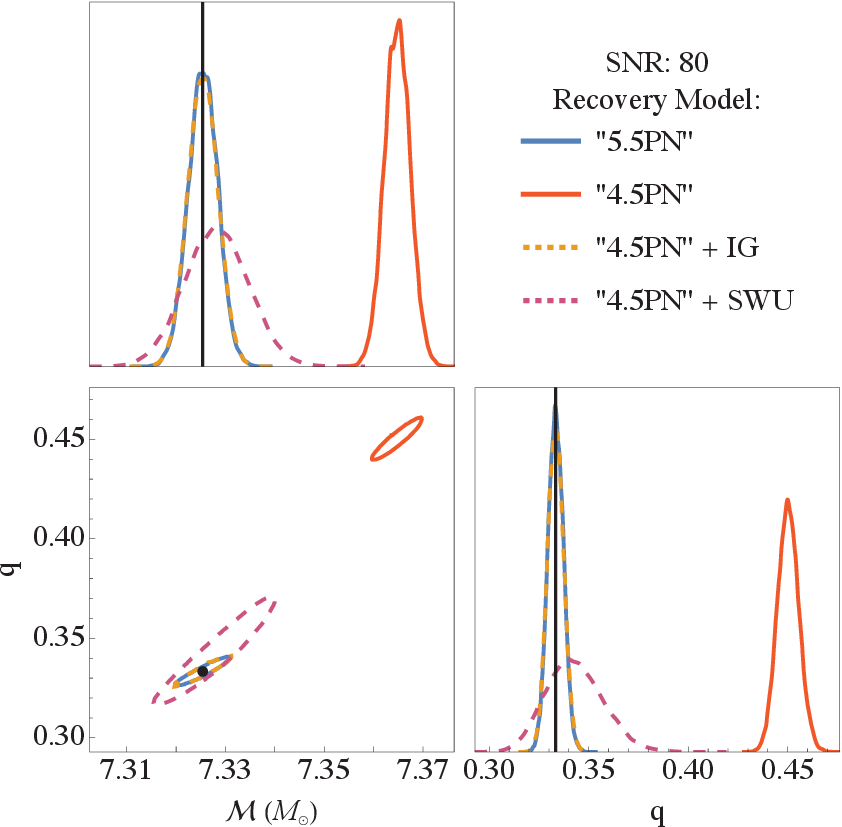}
\caption{\label{fig:margresults} Partial corner plot using the ``5.5PN'' and ``4.5PN'' recovery models, as well as the enhanced model with a ``4.5PN'' order base and the two priors (IG and SWU). All partial corner plots are marginalized over the parameters not shown in the figure. In particular, the enhanced ``4.5PN'' model are marginalized over the higher PN order terms introduced. The upper left and lower right panels contain the one-dimensional PDFs for chirp mass $\mathcal{M}$ and mass ratio $q$, respectively, and the lower left panel depicts the 90\% credible region contours, with the injection values are indicated in black. Observe that the enhanced models remove the systematic bias introduced by the plain ``4.5PN'' model, at the cost of enlarging the width of the posterior probability distribution when the range of $\bar{\sigma}_{3,4}$ is assumed unknown (the SWU prior).}
\end{figure}

With all of this in mind, let us now consider a particular example. 
Let us focus on the ``4.5PN'' recovery model, because the analyses of the previous section demonstrated that this model leads to the largest systematic biases relative to statistical error for both the equal and unequal mass injections.
Let us then enhance this ``4.5PN'' recovery model with two ignorance terms, one at ``5PN'' order (controlled by a new parameter $\bar{\sigma}_3$) and one at ``5.5PN'' order (controlled by a new parameter $\bar{\sigma}_4$). 
For the priors on $\bar{\sigma}_3$ and $\bar{\sigma}_4$, let us consider two cases:
\begin{itemize}
    \item \textbf{Informed Gaussian (IG):} We collect samples on $q$ produced during the recovery done with the ``5.5PN''  model. We then substitute these $q$ samples into the $\sigma_3(q)$ and $\sigma_4(q)$ functions (defined in the \texttt{IMRPhenomD} model through the fitted polynomial ansatz) to produce two-dimensional distributions in $(\sigma_3,\sigma_4)$. We use these distributions to fit a two-dimensional correlated Gaussian, and we use these Gaussian distributions as the priors on $\bar{\sigma}_3$ and $\bar{\sigma}_4$. 
    \item  \textbf{Super Wide Uniform (SWU):} With the prior bounds on $q \in [1/8,1]$, we infer upper and lower prior boundaries on $\bar{\sigma}_3$ and $\bar{\sigma}_4$ via $\bar{\sigma}_{3,4}^{\rm max/min} = \sigma_{3,4}(q_{\rm max/min})$. With this in hand, we then assign uniform and uncorrelated priors inside that range. 
\end{itemize}
While the IG prior is fully informed by the knowledge of the higher PN order terms, the SWU is much more relaxed. The way we have constructed the SWU prior does use knowledge of the higher PN order terms, but this is only a matter of convenience here. We could have, instead, used a completely agnostic approach by requiring that each $\bar{\sigma}_i$ term be smaller than the one preceding it. Such an assumption is applicable when considering inspirals, which can be modeled as PN series. We already demonstrated in Fig. \ref{FIG:allterms} that the terms in the phenomenological part of the {\tt IMRPhenomD} phase obey this requirement well. Therefore, the SWU prior ends up producing a prior that is very similar to what one would find with the agnostic approach just described. 

With this enhanced recovery model defined, we then carry out a Bayesian parameter estimation analysis for the unequal mass signal injection at SNR=80, the results of which are summarized in the partial corner plot of Fig. \ref{fig:margresults}.
The upper left and lower right panels represent the one-dimensional marginalized posterior probability distributions on $\mathcal{M}$ and $q$ respectively and the lower left panel depicts the 90\% credible region contours of the marginalized two-dimensional posterior probability distributions.
Let us compare the posterior probability distributions from the analyses done with the ``5.5PN'' recovery model, the standard ``4.5PN'' recovery model, and the enhanced ``4.5PN'' recovery model with the IG and SWU priors described above.
Observe that the enhanced recovery model removes the systematic bias of the ``4.5 PN'' model. 
Observe also that the use of the IG prior preserves the width of the posterior probability distribution, while the SWU increases this width, therefore increasing the amount of statistical uncertainty. 
This is as expected because the IG prior uses knowledge of the higher PN order terms that is not truly accessible, while the SWU prior models this ignorance at the cost of a wider posterior probability distribution. 
This analysis provides a proof-of-principle that the method proposed to marginalize over our ignorance in our models can produce much more accurate recovered parameters at the cost of more statistical error. 

%%%%%%%%%%%%%%%%%%%%%%%%%%%%%%%%%%%%%%%%%%%%%%%%%%%%%%%%%%%%%%%%%%%%%%%%%%%%%%%%%%%%%%%%%%
\section{Discussion}\label{SEC:Discussion}
In this paper, we have presented an injection and recovery campaign performed to understand how systematic error is impacted by PN corrections to the phase of the gravitational waves emitted in the quasi-circular inspiral portion of black hole binary coalescence.
We have considered injected data of equal and unequal mass, non-spinning black hole binaries for a selection of SNRs and recovered with models truncated at different PN orders in the {\texttt{IMRPhenomD}} phase. 
We have shown that even truncation at ``5PN'' order can lead to systematic error that dominates over statistical error at SNRs expected in the next observing runs of current detectors.
This does not necessarily indicate that the {\tt IMRPhenomD} model contains these errors, because the fits that produced the higher PN order terms of this model were done assuming the model would never be truncated at a given PN order.
However, our results demonstrate the importance of these higher-order corrections for accurate parameter estimation.

An important reminder that arises as a consequence of our work is that both the IMRPhenom and the EOB waveform models used in gravitational wave data analysis to date carry uncertainties even in the inspiral phase due to fits of certain (unknown high PN order) parameters. 
These fits are not perfect and carry both statistical error (due to the finite size of the ``data'' the models are fit to) and systematic error (due to the finite accuracy of the ``data'').  
To date, these errors are not accounted for in parameter estimation, and could, in principle, affect parameter inferences for sufficiently high SNR events. 
Our work suggests that this systematic error need not only be relevant to observations with third-generation detectors, but they could already influence parameter inferences in the next observing runs with current detectors. 

In an initial attempt to tackle this issue, we here propose a method to ameliorate the impact of systematic uncertainties in parameter estimation. 
Our proposal relies on the idea of creating a model to characterize our ignorance in the waveform and then to marginalize over it. 
In the inspiral phase, this can be done by promoting the fitting coefficients of IMRPhenom and EOB models to new waveform parameters (with priors set by the posterior probability distributions obtained in the fit) and by introducing higher PN order terms that capture terms not included in the waveform model. 
We have shown that this method can all but remove the systematic error introduced by truncating the inspiral phase at a fixed PN order, at the cost of widening the posterior probability distribution, and thus, increasing the statistical uncertainties.  

The work presented here opens several opportunities for further work. 
One such opportunity would be to investigate the inclusion of the fitting parameters of IMRPhenom and EOB models in parameter estimation, as described above. 
This would probably require re-doing the fits to numerical relativity simulations to account for (i) a larger set of such simulations and (ii) the finite accuracy of the simulations (with error estimates taken from the simulations themselves). 
Once these fits are re-done, the posterior probability distributions on the fitting parameters could be modeled through a multi-dimensional Gaussian or a kernel density estimator to prescribe priors for the fitting parameters in future parameter estimation studies. 

Another opportunity for future work would be to consider the impact of different (higher PN order) phenomenological functions when fitting against numerical relativity simulations in the inspiral. 
When developing IMR waveform models,
the degree of the polynomial to be fitted is determined by the information content of the numerical relativity signal and the inherent numerical error in the simulations. 
As the SNR of observed signals increases, as the error in numerical relativity simulations decreases, and as more widely applicable numerical relativity surrogate waveforms are constructed, the polynomial degree or even the functional form of the fitting functions in phenomenological waveforms could be revisited.
When constructing these models, emphasis could be placed on reducing bias in recovered parameters rather than increasing the fitting factor (or decreasing the mismatch) to numerical relativity simulations. 
The promising results of our proposed method to ameliorate systematic error presented here suggest that our ignorance of higher PN order terms (which will always exist, at any given finite PN order) is likely to not hinder future parameter inferences and the potential of gravitational waves to learn about astrophysics and fundamental physics.   

%%%%%%%%%%%%%%%%%%%%%%%%%%%%%%%%%%%%%%%%%%%%%%%%%%%%%%%%%%%%%%%%%%%%%%%%%%%%%%%%%%%%%%%%%%
\acknowledgements
%%%%%%%%%%%%%%%%%%%%%%%%%%%%%%%%%%%%%%%%%%%%%%%%%%%%%%%%%%%%%%%%%%%%%%%%%%%%%%%%%%%%%%%%%%
The authors are grateful to Katerina Chatziioannou, Jon Gair, Michael P\"urrer, Leo Stein,  Helvi Witek, and Simone Mezzasoma for useful comments and discussions. CBO and NY acknowledge support from NSF Grants PHY-1759615 and PHY-1949838. NJC acknowledges support from NSF Grants PHY-1912053 and PHY-2207970. The authors are also grateful for computational resources provided by the LIGO Laboratory and supported by National Science Foundation Grants PHY-0757058 and PHY-0823459.
This is LIGO Document Number LIGO-P2300015.

%%%%%%%%%%%%%%%%%%%%%%%%%%%%%%%%%%%%%%%%%%%%%%%%%%%%%%%%%%%%%%%%%%%%%%%%%%%%%%%%%%%%%%%%%%
%


\begin{thebibliography}{56}%
\makeatletter
\providecommand \@ifxundefined [1]{%
 \@ifx{#1\undefined}
}%
\providecommand \@ifnum [1]{%
 \ifnum #1\expandafter \@firstoftwo
 \else \expandafter \@secondoftwo
 \fi
}%
\providecommand \@ifx [1]{%
 \ifx #1\expandafter \@firstoftwo
 \else \expandafter \@secondoftwo
 \fi
}%
\providecommand \natexlab [1]{#1}%
\providecommand \enquote  [1]{``#1''}%
\providecommand \bibnamefont  [1]{#1}%
\providecommand \bibfnamefont [1]{#1}%
\providecommand \citenamefont [1]{#1}%
\providecommand \href@noop [0]{\@secondoftwo}%
\providecommand \href [0]{\begingroup \@sanitize@url \@href}%
\providecommand \@href[1]{\@@startlink{#1}\@@href}%
\providecommand \@@href[1]{\endgroup#1\@@endlink}%
\providecommand \@sanitize@url [0]{\catcode `\\12\catcode `\$12\catcode
  `\&12\catcode `\#12\catcode `\^12\catcode `\_12\catcode `\%12\relax}%
\providecommand \@@startlink[1]{}%
\providecommand \@@endlink[0]{}%
\providecommand \url  [0]{\begingroup\@sanitize@url \@url }%
\providecommand \@url [1]{\endgroup\@href {#1}{\urlprefix }}%
\providecommand \urlprefix  [0]{URL }%
\providecommand \Eprint [0]{\href }%
\providecommand \doibase [0]{http://dx.doi.org/}%
\providecommand \selectlanguage [0]{\@gobble}%
\providecommand \bibinfo  [0]{\@secondoftwo}%
\providecommand \bibfield  [0]{\@secondoftwo}%
\providecommand \translation [1]{[#1]}%
\providecommand \BibitemOpen [0]{}%
\providecommand \bibitemStop [0]{}%
\providecommand \bibitemNoStop [0]{.\EOS\space}%
\providecommand \EOS [0]{\spacefactor3000\relax}%
\providecommand \BibitemShut  [1]{\csname bibitem#1\endcsname}%
\let\auto@bib@innerbib\@empty
%</preamble>
\bibitem [{\citenamefont {Blanchet}(2002)}]{Blanchet:2002av}%
  \BibitemOpen
  \bibfield  {author} {\bibinfo {author} {\bibfnamefont {L.}~\bibnamefont
  {Blanchet}},\ }\href {\doibase 10.12942/lrr-2002-3} {\bibfield  {journal}
  {\bibinfo  {journal} {Living Rev. Rel.}\ }\textbf {\bibinfo {volume} {5}},\
  \bibinfo {pages} {3} (\bibinfo {year} {2002})},\ \Eprint
  {http://arxiv.org/abs/gr-qc/0202016} {arXiv:gr-qc/0202016} \BibitemShut
  {NoStop}%
\bibitem [{\citenamefont {Will}(2011)}]{Will:2011nz}%
  \BibitemOpen
  \bibfield  {author} {\bibinfo {author} {\bibfnamefont {C.~M.}\ \bibnamefont
  {Will}},\ }\href {\doibase 10.1073/pnas.1103127108} {\bibfield  {journal}
  {\bibinfo  {journal} {Proc. Nat. Acad. Sci.}\ }\textbf {\bibinfo {volume}
  {108}},\ \bibinfo {pages} {5938} (\bibinfo {year} {2011})},\ \Eprint
  {http://arxiv.org/abs/1102.5192} {arXiv:1102.5192 [gr-qc]} \BibitemShut
  {NoStop}%
\bibitem [{\citenamefont {Buonanno}\ and\ \citenamefont
  {Damour}(1999)}]{Buonanno:1998gg}%
  \BibitemOpen
  \bibfield  {author} {\bibinfo {author} {\bibfnamefont {A.}~\bibnamefont
  {Buonanno}}\ and\ \bibinfo {author} {\bibfnamefont {T.}~\bibnamefont
  {Damour}},\ }\href {\doibase 10.1103/PhysRevD.59.084006} {\bibfield
  {journal} {\bibinfo  {journal} {Phys. Rev. D}\ }\textbf {\bibinfo {volume}
  {59}},\ \bibinfo {pages} {084006} (\bibinfo {year} {1999})},\ \Eprint
  {http://arxiv.org/abs/gr-qc/9811091} {arXiv:gr-qc/9811091} \BibitemShut
  {NoStop}%
\bibitem [{\citenamefont {Boh\'e}\ \emph {et~al.}(2017)\citenamefont {Boh\'e}
  \emph {et~al.}}]{Bohe:2016gbl}%
  \BibitemOpen
  \bibfield  {author} {\bibinfo {author} {\bibfnamefont {A.}~\bibnamefont
  {Boh\'e}} \emph {et~al.},\ }\href {\doibase 10.1103/PhysRevD.95.044028}
  {\bibfield  {journal} {\bibinfo  {journal} {Phys. Rev. D}\ }\textbf {\bibinfo
  {volume} {95}},\ \bibinfo {pages} {044028} (\bibinfo {year} {2017})},\
  \Eprint {http://arxiv.org/abs/1611.03703} {arXiv:1611.03703 [gr-qc]}
  \BibitemShut {NoStop}%
\bibitem [{\citenamefont {Cotesta}\ \emph {et~al.}(2018)\citenamefont
  {Cotesta}, \citenamefont {Buonanno}, \citenamefont {Boh\'e}, \citenamefont
  {Taracchini}, \citenamefont {Hinder},\ and\ \citenamefont
  {Ossokine}}]{Cotesta:2018fcv}%
  \BibitemOpen
  \bibfield  {author} {\bibinfo {author} {\bibfnamefont {R.}~\bibnamefont
  {Cotesta}}, \bibinfo {author} {\bibfnamefont {A.}~\bibnamefont {Buonanno}},
  \bibinfo {author} {\bibfnamefont {A.}~\bibnamefont {Boh\'e}}, \bibinfo
  {author} {\bibfnamefont {A.}~\bibnamefont {Taracchini}}, \bibinfo {author}
  {\bibfnamefont {I.}~\bibnamefont {Hinder}}, \ and\ \bibinfo {author}
  {\bibfnamefont {S.}~\bibnamefont {Ossokine}},\ }\href {\doibase
  10.1103/PhysRevD.98.084028} {\bibfield  {journal} {\bibinfo  {journal} {Phys.
  Rev. D}\ }\textbf {\bibinfo {volume} {98}},\ \bibinfo {pages} {084028}
  (\bibinfo {year} {2018})},\ \Eprint {http://arxiv.org/abs/1803.10701}
  {arXiv:1803.10701 [gr-qc]} \BibitemShut {NoStop}%
\bibitem [{\citenamefont {Nagar}\ \emph {et~al.}(2018)\citenamefont {Nagar}
  \emph {et~al.}}]{Nagar:2018zoe}%
  \BibitemOpen
  \bibfield  {author} {\bibinfo {author} {\bibfnamefont {A.}~\bibnamefont
  {Nagar}} \emph {et~al.},\ }\href {\doibase 10.1103/PhysRevD.98.104052}
  {\bibfield  {journal} {\bibinfo  {journal} {Phys. Rev. D}\ }\textbf {\bibinfo
  {volume} {98}},\ \bibinfo {pages} {104052} (\bibinfo {year} {2018})},\
  \Eprint {http://arxiv.org/abs/1806.01772} {arXiv:1806.01772 [gr-qc]}
  \BibitemShut {NoStop}%
\bibitem [{\citenamefont {Nagar}\ \emph {et~al.}(2020)\citenamefont {Nagar},
  \citenamefont {Riemenschneider}, \citenamefont {Pratten}, \citenamefont
  {Rettegno},\ and\ \citenamefont {Messina}}]{Nagar:2020pcj}%
  \BibitemOpen
  \bibfield  {author} {\bibinfo {author} {\bibfnamefont {A.}~\bibnamefont
  {Nagar}}, \bibinfo {author} {\bibfnamefont {G.}~\bibnamefont
  {Riemenschneider}}, \bibinfo {author} {\bibfnamefont {G.}~\bibnamefont
  {Pratten}}, \bibinfo {author} {\bibfnamefont {P.}~\bibnamefont {Rettegno}}, \
  and\ \bibinfo {author} {\bibfnamefont {F.}~\bibnamefont {Messina}},\ }\href
  {\doibase 10.1103/PhysRevD.102.024077} {\bibfield  {journal} {\bibinfo
  {journal} {Phys. Rev. D}\ }\textbf {\bibinfo {volume} {102}},\ \bibinfo
  {pages} {024077} (\bibinfo {year} {2020})},\ \Eprint
  {http://arxiv.org/abs/2001.09082} {arXiv:2001.09082 [gr-qc]} \BibitemShut
  {NoStop}%
\bibitem [{\citenamefont {Ossokine}\ \emph {et~al.}(2020)\citenamefont
  {Ossokine} \emph {et~al.}}]{Ossokine:2020kjp}%
  \BibitemOpen
  \bibfield  {author} {\bibinfo {author} {\bibfnamefont {S.}~\bibnamefont
  {Ossokine}} \emph {et~al.},\ }\href {\doibase 10.1103/PhysRevD.102.044055}
  {\bibfield  {journal} {\bibinfo  {journal} {Phys. Rev. D}\ }\textbf {\bibinfo
  {volume} {102}},\ \bibinfo {pages} {044055} (\bibinfo {year} {2020})},\
  \Eprint {http://arxiv.org/abs/2004.09442} {arXiv:2004.09442 [gr-qc]}
  \BibitemShut {NoStop}%
\bibitem [{\citenamefont {Husa}\ \emph {et~al.}(2016)\citenamefont {Husa},
  \citenamefont {Khan}, \citenamefont {Hannam}, \citenamefont {P\"urrer},
  \citenamefont {Ohme}, \citenamefont {Jim\'enez~Forteza},\ and\ \citenamefont
  {Boh\'e}}]{Husa:2015iqa}%
  \BibitemOpen
  \bibfield  {author} {\bibinfo {author} {\bibfnamefont {S.}~\bibnamefont
  {Husa}}, \bibinfo {author} {\bibfnamefont {S.}~\bibnamefont {Khan}}, \bibinfo
  {author} {\bibfnamefont {M.}~\bibnamefont {Hannam}}, \bibinfo {author}
  {\bibfnamefont {M.}~\bibnamefont {P\"urrer}}, \bibinfo {author}
  {\bibfnamefont {F.}~\bibnamefont {Ohme}}, \bibinfo {author} {\bibfnamefont
  {X.}~\bibnamefont {Jim\'enez~Forteza}}, \ and\ \bibinfo {author}
  {\bibfnamefont {A.}~\bibnamefont {Boh\'e}},\ }\href {\doibase
  10.1103/PhysRevD.93.044006} {\bibfield  {journal} {\bibinfo  {journal} {Phys.
  Rev. D}\ }\textbf {\bibinfo {volume} {93}},\ \bibinfo {pages} {044006}
  (\bibinfo {year} {2016})},\ \Eprint {http://arxiv.org/abs/1508.07250}
  {arXiv:1508.07250 [gr-qc]} \BibitemShut {NoStop}%
\bibitem [{\citenamefont {Khan}\ \emph {et~al.}(2016)\citenamefont {Khan},
  \citenamefont {Husa}, \citenamefont {Hannam}, \citenamefont {Ohme},
  \citenamefont {P\"urrer}, \citenamefont {Jim\'enez~Forteza},\ and\
  \citenamefont {Boh\'e}}]{Khan:2015jqa}%
  \BibitemOpen
  \bibfield  {author} {\bibinfo {author} {\bibfnamefont {S.}~\bibnamefont
  {Khan}}, \bibinfo {author} {\bibfnamefont {S.}~\bibnamefont {Husa}}, \bibinfo
  {author} {\bibfnamefont {M.}~\bibnamefont {Hannam}}, \bibinfo {author}
  {\bibfnamefont {F.}~\bibnamefont {Ohme}}, \bibinfo {author} {\bibfnamefont
  {M.}~\bibnamefont {P\"urrer}}, \bibinfo {author} {\bibfnamefont
  {X.}~\bibnamefont {Jim\'enez~Forteza}}, \ and\ \bibinfo {author}
  {\bibfnamefont {A.}~\bibnamefont {Boh\'e}},\ }\href {\doibase
  10.1103/PhysRevD.93.044007} {\bibfield  {journal} {\bibinfo  {journal} {Phys.
  Rev. D}\ }\textbf {\bibinfo {volume} {93}},\ \bibinfo {pages} {044007}
  (\bibinfo {year} {2016})},\ \Eprint {http://arxiv.org/abs/1508.07253}
  {arXiv:1508.07253 [gr-qc]} \BibitemShut {NoStop}%
\bibitem [{\citenamefont {Pratten}\ \emph {et~al.}(2020)\citenamefont
  {Pratten}, \citenamefont {Husa}, \citenamefont {Garcia-Quiros}, \citenamefont
  {Colleoni}, \citenamefont {Ramos-Buades}, \citenamefont {Estelles},\ and\
  \citenamefont {Jaume}}]{Pratten:2020fqn}%
  \BibitemOpen
  \bibfield  {author} {\bibinfo {author} {\bibfnamefont {G.}~\bibnamefont
  {Pratten}}, \bibinfo {author} {\bibfnamefont {S.}~\bibnamefont {Husa}},
  \bibinfo {author} {\bibfnamefont {C.}~\bibnamefont {Garcia-Quiros}}, \bibinfo
  {author} {\bibfnamefont {M.}~\bibnamefont {Colleoni}}, \bibinfo {author}
  {\bibfnamefont {A.}~\bibnamefont {Ramos-Buades}}, \bibinfo {author}
  {\bibfnamefont {H.}~\bibnamefont {Estelles}}, \ and\ \bibinfo {author}
  {\bibfnamefont {R.}~\bibnamefont {Jaume}},\ }\href {\doibase
  10.1103/PhysRevD.102.064001} {\bibfield  {journal} {\bibinfo  {journal}
  {Phys. Rev. D}\ }\textbf {\bibinfo {volume} {102}},\ \bibinfo {pages}
  {064001} (\bibinfo {year} {2020})},\ \Eprint
  {http://arxiv.org/abs/2001.11412} {arXiv:2001.11412 [gr-qc]} \BibitemShut
  {NoStop}%
\bibitem [{\citenamefont {Pratten}\ \emph {et~al.}(2021)\citenamefont {Pratten}
  \emph {et~al.}}]{Pratten:2020ceb}%
  \BibitemOpen
  \bibfield  {author} {\bibinfo {author} {\bibfnamefont {G.}~\bibnamefont
  {Pratten}} \emph {et~al.},\ }\href {\doibase 10.1103/PhysRevD.103.104056}
  {\bibfield  {journal} {\bibinfo  {journal} {Phys. Rev. D}\ }\textbf {\bibinfo
  {volume} {103}},\ \bibinfo {pages} {104056} (\bibinfo {year} {2021})},\
  \Eprint {http://arxiv.org/abs/2004.06503} {arXiv:2004.06503 [gr-qc]}
  \BibitemShut {NoStop}%
\bibitem [{\citenamefont {Hinder}\ \emph {et~al.}(2014)\citenamefont {Hinder}
  \emph {et~al.}}]{Hinder:2013oqa}%
  \BibitemOpen
  \bibfield  {author} {\bibinfo {author} {\bibfnamefont {I.}~\bibnamefont
  {Hinder}} \emph {et~al.},\ }\href {\doibase 10.1088/0264-9381/31/2/025012}
  {\bibfield  {journal} {\bibinfo  {journal} {Class. Quant. Grav.}\ }\textbf
  {\bibinfo {volume} {31}},\ \bibinfo {pages} {025012} (\bibinfo {year}
  {2014})},\ \Eprint {http://arxiv.org/abs/1307.5307} {arXiv:1307.5307 [gr-qc]}
  \BibitemShut {NoStop}%
\bibitem [{\citenamefont {Aasi}\ \emph {et~al.}(2015)\citenamefont {Aasi} \emph
  {et~al.}}]{LIGOScientific:2014pky}%
  \BibitemOpen
  \bibfield  {author} {\bibinfo {author} {\bibfnamefont {J.}~\bibnamefont
  {Aasi}} \emph {et~al.} (\bibinfo {collaboration} {LIGO Scientific}),\ }\href
  {\doibase 10.1088/0264-9381/32/7/074001} {\bibfield  {journal} {\bibinfo
  {journal} {Class. Quant. Grav.}\ }\textbf {\bibinfo {volume} {32}},\ \bibinfo
  {pages} {074001} (\bibinfo {year} {2015})},\ \Eprint
  {http://arxiv.org/abs/1411.4547} {arXiv:1411.4547 [gr-qc]} \BibitemShut
  {NoStop}%
\bibitem [{\citenamefont {Acernese}\ \emph {et~al.}(2015)\citenamefont
  {Acernese} \emph {et~al.}}]{TheVirgo:2014hva}%
  \BibitemOpen
  \bibfield  {author} {\bibinfo {author} {\bibfnamefont {F.}~\bibnamefont
  {Acernese}} \emph {et~al.} (\bibinfo {collaboration} {VIRGO}),\ }\href
  {\doibase 10.1088/0264-9381/32/2/024001} {\bibfield  {journal} {\bibinfo
  {journal} {Class. Quant. Grav.}\ }\textbf {\bibinfo {volume} {32}},\ \bibinfo
  {pages} {024001} (\bibinfo {year} {2015})},\ \Eprint
  {http://arxiv.org/abs/1408.3978} {arXiv:1408.3978 [gr-qc]} \BibitemShut
  {NoStop}%
\bibitem [{\citenamefont {Abbott}\ \emph
  {et~al.}(2021{\natexlab{a}})\citenamefont {Abbott} \emph
  {et~al.}}]{LIGOScientific:2021usb}%
  \BibitemOpen
  \bibfield  {author} {\bibinfo {author} {\bibfnamefont {R.}~\bibnamefont
  {Abbott}} \emph {et~al.} (\bibinfo {collaboration} {LIGO Scientific,
  VIRGO}),\ }\href@noop {} {\  (\bibinfo {year} {2021}{\natexlab{a}})},\
  \Eprint {http://arxiv.org/abs/2108.01045} {arXiv:2108.01045 [gr-qc]}
  \BibitemShut {NoStop}%
\bibitem [{\citenamefont {Abbott}\ \emph
  {et~al.}(2021{\natexlab{b}})\citenamefont {Abbott} \emph
  {et~al.}}]{LIGOScientific:2021djp}%
  \BibitemOpen
  \bibfield  {author} {\bibinfo {author} {\bibfnamefont {R.}~\bibnamefont
  {Abbott}} \emph {et~al.} (\bibinfo {collaboration} {LIGO Scientific, VIRGO,
  KAGRA}),\ }\href@noop {} {\  (\bibinfo {year} {2021}{\natexlab{b}})},\
  \Eprint {http://arxiv.org/abs/2111.03606} {arXiv:2111.03606 [gr-qc]}
  \BibitemShut {NoStop}%
\bibitem [{\citenamefont {Varma}\ \emph {et~al.}(2019)\citenamefont {Varma},
  \citenamefont {Field}, \citenamefont {Scheel}, \citenamefont {Blackman},
  \citenamefont {Gerosa}, \citenamefont {Stein}, \citenamefont {Kidder},\ and\
  \citenamefont {Pfeiffer}}]{Varma:2019csw}%
  \BibitemOpen
  \bibfield  {author} {\bibinfo {author} {\bibfnamefont {V.}~\bibnamefont
  {Varma}}, \bibinfo {author} {\bibfnamefont {S.~E.}\ \bibnamefont {Field}},
  \bibinfo {author} {\bibfnamefont {M.~A.}\ \bibnamefont {Scheel}}, \bibinfo
  {author} {\bibfnamefont {J.}~\bibnamefont {Blackman}}, \bibinfo {author}
  {\bibfnamefont {D.}~\bibnamefont {Gerosa}}, \bibinfo {author} {\bibfnamefont
  {L.~C.}\ \bibnamefont {Stein}}, \bibinfo {author} {\bibfnamefont {L.~E.}\
  \bibnamefont {Kidder}}, \ and\ \bibinfo {author} {\bibfnamefont {H.~P.}\
  \bibnamefont {Pfeiffer}},\ }\href {\doibase 10.1103/PhysRevResearch.1.033015}
  {\bibfield  {journal} {\bibinfo  {journal} {Phys. Rev. Research.}\ }\textbf
  {\bibinfo {volume} {1}},\ \bibinfo {pages} {033015} (\bibinfo {year}
  {2019})},\ \Eprint {http://arxiv.org/abs/1905.09300} {arXiv:1905.09300
  [gr-qc]} \BibitemShut {NoStop}%
\bibitem [{\citenamefont {Islam}\ \emph {et~al.}(2022)\citenamefont {Islam},
  \citenamefont {Field}, \citenamefont {Hughes}, \citenamefont {Khanna},
  \citenamefont {Varma}, \citenamefont {Giesler}, \citenamefont {Scheel},
  \citenamefont {Kidder},\ and\ \citenamefont {Pfeiffer}}]{Islam:2022laz}%
  \BibitemOpen
  \bibfield  {author} {\bibinfo {author} {\bibfnamefont {T.}~\bibnamefont
  {Islam}}, \bibinfo {author} {\bibfnamefont {S.~E.}\ \bibnamefont {Field}},
  \bibinfo {author} {\bibfnamefont {S.~A.}\ \bibnamefont {Hughes}}, \bibinfo
  {author} {\bibfnamefont {G.}~\bibnamefont {Khanna}}, \bibinfo {author}
  {\bibfnamefont {V.}~\bibnamefont {Varma}}, \bibinfo {author} {\bibfnamefont
  {M.}~\bibnamefont {Giesler}}, \bibinfo {author} {\bibfnamefont {M.~A.}\
  \bibnamefont {Scheel}}, \bibinfo {author} {\bibfnamefont {L.~E.}\
  \bibnamefont {Kidder}}, \ and\ \bibinfo {author} {\bibfnamefont {H.~P.}\
  \bibnamefont {Pfeiffer}},\ }\href {\doibase 10.1103/PhysRevD.106.104025}
  {\bibfield  {journal} {\bibinfo  {journal} {Phys. Rev. D}\ }\textbf {\bibinfo
  {volume} {106}},\ \bibinfo {pages} {104025} (\bibinfo {year} {2022})},\
  \Eprint {http://arxiv.org/abs/2204.01972} {arXiv:2204.01972 [gr-qc]}
  \BibitemShut {NoStop}%
\bibitem [{\citenamefont {Cutler}\ and\ \citenamefont
  {Vallisneri}(2007)}]{Cutler:2007mi}%
  \BibitemOpen
  \bibfield  {author} {\bibinfo {author} {\bibfnamefont {C.}~\bibnamefont
  {Cutler}}\ and\ \bibinfo {author} {\bibfnamefont {M.}~\bibnamefont
  {Vallisneri}},\ }\href {\doibase 10.1103/PhysRevD.76.104018} {\bibfield
  {journal} {\bibinfo  {journal} {Phys. Rev. D}\ }\textbf {\bibinfo {volume}
  {76}},\ \bibinfo {pages} {104018} (\bibinfo {year} {2007})},\ \Eprint
  {http://arxiv.org/abs/0707.2982} {arXiv:0707.2982 [gr-qc]} \BibitemShut
  {NoStop}%
\bibitem [{\citenamefont {Canitrot}(2001)}]{Canitrot:2001hc}%
  \BibitemOpen
  \bibfield  {author} {\bibinfo {author} {\bibfnamefont {P.}~\bibnamefont
  {Canitrot}},\ }\href {\doibase 10.1103/PhysRevD.63.082005} {\bibfield
  {journal} {\bibinfo  {journal} {Phys. Rev. D}\ }\textbf {\bibinfo {volume}
  {63}},\ \bibinfo {pages} {082005} (\bibinfo {year} {2001})}\BibitemShut
  {NoStop}%
\bibitem [{\citenamefont {Moore}\ \emph {et~al.}(2021)\citenamefont {Moore},
  \citenamefont {Finch}, \citenamefont {Buscicchio},\ and\ \citenamefont
  {Gerosa}}]{Moore:2021eok}%
  \BibitemOpen
  \bibfield  {author} {\bibinfo {author} {\bibfnamefont {C.~J.}\ \bibnamefont
  {Moore}}, \bibinfo {author} {\bibfnamefont {E.}~\bibnamefont {Finch}},
  \bibinfo {author} {\bibfnamefont {R.}~\bibnamefont {Buscicchio}}, \ and\
  \bibinfo {author} {\bibfnamefont {D.}~\bibnamefont {Gerosa}},\ }\href
  {\doibase 10.1016/j.isci.2021.102577} {\  (\bibinfo {year} {2021}),\
  10.1016/j.isci.2021.102577},\ \Eprint {http://arxiv.org/abs/2103.16486}
  {arXiv:2103.16486 [gr-qc]} \BibitemShut {NoStop}%
\bibitem [{\citenamefont {Littenberg}\ \emph {et~al.}(2013)\citenamefont
  {Littenberg}, \citenamefont {Baker}, \citenamefont {Buonanno},\ and\
  \citenamefont {Kelly}}]{Littenberg:2012uj}%
  \BibitemOpen
  \bibfield  {author} {\bibinfo {author} {\bibfnamefont {T.~B.}\ \bibnamefont
  {Littenberg}}, \bibinfo {author} {\bibfnamefont {J.~G.}\ \bibnamefont
  {Baker}}, \bibinfo {author} {\bibfnamefont {A.}~\bibnamefont {Buonanno}}, \
  and\ \bibinfo {author} {\bibfnamefont {B.~J.}\ \bibnamefont {Kelly}},\ }\href
  {\doibase 10.1103/PhysRevD.87.104003} {\bibfield  {journal} {\bibinfo
  {journal} {Phys. Rev. D}\ }\textbf {\bibinfo {volume} {87}},\ \bibinfo
  {pages} {104003} (\bibinfo {year} {2013})},\ \Eprint
  {http://arxiv.org/abs/1210.0893} {arXiv:1210.0893 [gr-qc]} \BibitemShut
  {NoStop}%
\bibitem [{\citenamefont {P\"urrer}\ and\ \citenamefont
  {Haster}(2020)}]{Purrer:2019jcp}%
  \BibitemOpen
  \bibfield  {author} {\bibinfo {author} {\bibfnamefont {M.}~\bibnamefont
  {P\"urrer}}\ and\ \bibinfo {author} {\bibfnamefont {C.-J.}\ \bibnamefont
  {Haster}},\ }\href {\doibase 10.1103/PhysRevResearch.2.023151} {\bibfield
  {journal} {\bibinfo  {journal} {Phys. Rev. Res.}\ }\textbf {\bibinfo {volume}
  {2}},\ \bibinfo {pages} {023151} (\bibinfo {year} {2020})},\ \Eprint
  {http://arxiv.org/abs/1912.10055} {arXiv:1912.10055 [gr-qc]} \BibitemShut
  {NoStop}%
\bibitem [{\citenamefont {Chua}\ \emph {et~al.}(2020)\citenamefont {Chua},
  \citenamefont {Korsakova}, \citenamefont {Moore}, \citenamefont {Gair},\ and\
  \citenamefont {Babak}}]{Chua:2019wgs}%
  \BibitemOpen
  \bibfield  {author} {\bibinfo {author} {\bibfnamefont {A.~J.~K.}\
  \bibnamefont {Chua}}, \bibinfo {author} {\bibfnamefont {N.}~\bibnamefont
  {Korsakova}}, \bibinfo {author} {\bibfnamefont {C.~J.}\ \bibnamefont
  {Moore}}, \bibinfo {author} {\bibfnamefont {J.~R.}\ \bibnamefont {Gair}}, \
  and\ \bibinfo {author} {\bibfnamefont {S.}~\bibnamefont {Babak}},\ }\href
  {\doibase 10.1103/PhysRevD.101.044027} {\bibfield  {journal} {\bibinfo
  {journal} {Phys. Rev. D}\ }\textbf {\bibinfo {volume} {101}},\ \bibinfo
  {pages} {044027} (\bibinfo {year} {2020})},\ \Eprint
  {http://arxiv.org/abs/1912.11543} {arXiv:1912.11543 [astro-ph.IM]}
  \BibitemShut {NoStop}%
\bibitem [{\citenamefont {Vallisneri}(2008)}]{Vallisneri:2007ev}%
  \BibitemOpen
  \bibfield  {author} {\bibinfo {author} {\bibfnamefont {M.}~\bibnamefont
  {Vallisneri}},\ }\href {\doibase 10.1103/PhysRevD.77.042001} {\bibfield
  {journal} {\bibinfo  {journal} {Phys. Rev. D}\ }\textbf {\bibinfo {volume}
  {77}},\ \bibinfo {pages} {042001} (\bibinfo {year} {2008})},\ \Eprint
  {http://arxiv.org/abs/gr-qc/0703086} {arXiv:gr-qc/0703086} \BibitemShut
  {NoStop}%
\bibitem [{\citenamefont {Amaro-Seoane}\ \emph {et~al.}(2017)\citenamefont
  {Amaro-Seoane} \emph {et~al.}}]{amaroseoane2017laser}%
  \BibitemOpen
  \bibfield  {author} {\bibinfo {author} {\bibfnamefont {P.}~\bibnamefont
  {Amaro-Seoane}} \emph {et~al.},\ }\href@noop {} {\  (\bibinfo {year}
  {2017})},\ \Eprint {http://arxiv.org/abs/1702.00786} {arXiv:1702.00786}
  \BibitemShut {NoStop}%
\bibitem [{\citenamefont {Kaiser}\ and\ \citenamefont
  {McWilliams}(2021)}]{Kaiser:2020tlg}%
  \BibitemOpen
  \bibfield  {author} {\bibinfo {author} {\bibfnamefont {A.~R.}\ \bibnamefont
  {Kaiser}}\ and\ \bibinfo {author} {\bibfnamefont {S.~T.}\ \bibnamefont
  {McWilliams}},\ }\href {\doibase 10.1088/1361-6382/abd4f6} {\bibfield
  {journal} {\bibinfo  {journal} {Class. Quant. Grav.}\ }\textbf {\bibinfo
  {volume} {38}},\ \bibinfo {pages} {055009} (\bibinfo {year} {2021})},\
  \Eprint {http://arxiv.org/abs/2010.02135} {arXiv:2010.02135 [gr-qc]}
  \BibitemShut {NoStop}%
\bibitem [{\citenamefont {Abbott}\ \emph {et~al.}(2020)\citenamefont {Abbott}
  \emph {et~al.}}]{LIGOScientific:2020ufj}%
  \BibitemOpen
  \bibfield  {author} {\bibinfo {author} {\bibfnamefont {R.}~\bibnamefont
  {Abbott}} \emph {et~al.} (\bibinfo {collaboration} {LIGO Scientific,
  Virgo}),\ }\href {\doibase 10.3847/2041-8213/aba493} {\bibfield  {journal}
  {\bibinfo  {journal} {Astrophys. J. Lett.}\ }\textbf {\bibinfo {volume}
  {900}},\ \bibinfo {pages} {L13} (\bibinfo {year} {2020})},\ \Eprint
  {http://arxiv.org/abs/2009.01190} {arXiv:2009.01190 [astro-ph.HE]}
  \BibitemShut {NoStop}%
\bibitem [{\citenamefont {Abbott}\ \emph {et~al.}(2017)\citenamefont {Abbott}
  \emph {et~al.}}]{LIGOScientific:2016ebw}%
  \BibitemOpen
  \bibfield  {author} {\bibinfo {author} {\bibfnamefont {B.~P.}\ \bibnamefont
  {Abbott}} \emph {et~al.} (\bibinfo {collaboration} {LIGO Scientific,
  Virgo}),\ }\href {\doibase 10.1088/1361-6382/aa6854} {\bibfield  {journal}
  {\bibinfo  {journal} {Class. Quant. Grav.}\ }\textbf {\bibinfo {volume}
  {34}},\ \bibinfo {pages} {104002} (\bibinfo {year} {2017})},\ \Eprint
  {http://arxiv.org/abs/1611.07531} {arXiv:1611.07531 [gr-qc]} \BibitemShut
  {NoStop}%
\bibitem [{\citenamefont {Islam}\ \emph {et~al.}(2021)\citenamefont {Islam},
  \citenamefont {Field}, \citenamefont {Haster},\ and\ \citenamefont
  {Smith}}]{Islam:2020reh}%
  \BibitemOpen
  \bibfield  {author} {\bibinfo {author} {\bibfnamefont {T.}~\bibnamefont
  {Islam}}, \bibinfo {author} {\bibfnamefont {S.~E.}\ \bibnamefont {Field}},
  \bibinfo {author} {\bibfnamefont {C.-J.}\ \bibnamefont {Haster}}, \ and\
  \bibinfo {author} {\bibfnamefont {R.}~\bibnamefont {Smith}},\ }\href
  {\doibase 10.1103/PhysRevD.103.104027} {\bibfield  {journal} {\bibinfo
  {journal} {Phys. Rev. D}\ }\textbf {\bibinfo {volume} {103}},\ \bibinfo
  {pages} {104027} (\bibinfo {year} {2021})},\ \Eprint
  {http://arxiv.org/abs/2010.04848} {arXiv:2010.04848 [gr-qc]} \BibitemShut
  {NoStop}%
\bibitem [{\citenamefont {Abbott}\ \emph
  {et~al.}(2018{\natexlab{a}})\citenamefont {Abbott} \emph
  {et~al.}}]{KAGRA:2013rdx}%
  \BibitemOpen
  \bibfield  {author} {\bibinfo {author} {\bibfnamefont {B.~P.}\ \bibnamefont
  {Abbott}} \emph {et~al.} (\bibinfo {collaboration} {KAGRA, LIGO Scientific,
  Virgo, VIRGO}),\ }\href {\doibase 10.1007/s41114-020-00026-9} {\bibfield
  {journal} {\bibinfo  {journal} {Living Rev. Rel.}\ }\textbf {\bibinfo
  {volume} {21}},\ \bibinfo {pages} {3} (\bibinfo {year}
  {2018}{\natexlab{a}})},\ \Eprint {http://arxiv.org/abs/1304.0670}
  {arXiv:1304.0670 [gr-qc]} \BibitemShut {NoStop}%
\bibitem [{\citenamefont {Damour}\ \emph {et~al.}(2002)\citenamefont {Damour},
  \citenamefont {Iyer},\ and\ \citenamefont {Sathyaprakash}}]{Damour:2002kr}%
  \BibitemOpen
  \bibfield  {author} {\bibinfo {author} {\bibfnamefont {T.}~\bibnamefont
  {Damour}}, \bibinfo {author} {\bibfnamefont {B.~R.}\ \bibnamefont {Iyer}}, \
  and\ \bibinfo {author} {\bibfnamefont {B.~S.}\ \bibnamefont
  {Sathyaprakash}},\ }\href {\doibase 10.1103/PhysRevD.66.027502} {\bibfield
  {journal} {\bibinfo  {journal} {Phys. Rev. D}\ }\textbf {\bibinfo {volume}
  {66}},\ \bibinfo {pages} {027502} (\bibinfo {year} {2002})},\ \Eprint
  {http://arxiv.org/abs/gr-qc/0207021} {arXiv:gr-qc/0207021} \BibitemShut
  {NoStop}%
\bibitem [{\citenamefont {O'Reilly}\ \emph {et~al.}(2022)\citenamefont
  {O'Reilly}, \citenamefont {Branchesi},\ and\ \citenamefont
  {Gemme}}]{OReilly_2022}%
  \BibitemOpen
  \bibfield  {author} {\bibinfo {author} {\bibfnamefont {B.}~\bibnamefont
  {O'Reilly}}, \bibinfo {author} {\bibfnamefont {M.}~\bibnamefont {Branchesi}},
  \ and\ \bibinfo {author} {\bibfnamefont {G.}~\bibnamefont {Gemme}},\
  }\href@noop {} {\emph {\bibinfo {title} {LIGO Technical Document
  T2000012-v2}}},\ \bibinfo {type} {Tech. Rep.}\ (\bibinfo {year}
  {2022})\BibitemShut {NoStop}%
\bibitem [{\citenamefont {Ashton}\ \emph {et~al.}(2019)\citenamefont {Ashton}
  \emph {et~al.}}]{Ashton:2018jfp}%
  \BibitemOpen
  \bibfield  {author} {\bibinfo {author} {\bibfnamefont {G.}~\bibnamefont
  {Ashton}} \emph {et~al.},\ }\href {\doibase 10.3847/1538-4365/ab06fc}
  {\bibfield  {journal} {\bibinfo  {journal} {Astrophys. J. Suppl.}\ }\textbf
  {\bibinfo {volume} {241}},\ \bibinfo {pages} {27} (\bibinfo {year} {2019})},\
  \Eprint {http://arxiv.org/abs/1811.02042} {arXiv:1811.02042 [astro-ph.IM]}
  \BibitemShut {NoStop}%
\bibitem [{\citenamefont {Romero-Shaw}\ \emph {et~al.}(2020)\citenamefont
  {Romero-Shaw} \emph {et~al.}}]{Romero-Shaw:2020owr}%
  \BibitemOpen
  \bibfield  {author} {\bibinfo {author} {\bibfnamefont {I.~M.}\ \bibnamefont
  {Romero-Shaw}} \emph {et~al.},\ }\href {\doibase 10.1093/mnras/staa2850}
  {\bibfield  {journal} {\bibinfo  {journal} {Mon. Not. Roy. Astron. Soc.}\
  }\textbf {\bibinfo {volume} {499}},\ \bibinfo {pages} {3295} (\bibinfo {year}
  {2020})},\ \Eprint {http://arxiv.org/abs/2006.00714} {arXiv:2006.00714
  [astro-ph.IM]} \BibitemShut {NoStop}%
\bibitem [{\citenamefont {{LIGO Scientific Collaboration}}(2018)}]{lalsuite}%
  \BibitemOpen
  \bibfield  {author} {\bibinfo {author} {\bibnamefont {{LIGO Scientific
  Collaboration}}},\ }\href {\doibase 10.7935/GT1W-FZ16} {\enquote {\bibinfo
  {title} {{LIGO} {A}lgorithm {L}ibrary - {LALS}uite},}\ }\bibinfo
  {howpublished} {free software (GPL)} (\bibinfo {year} {2018})\BibitemShut
  {NoStop}%
\bibitem [{\citenamefont {Speagle}(2020)}]{Speagle_2020}%
  \BibitemOpen
  \bibfield  {author} {\bibinfo {author} {\bibfnamefont {J.~S.}\ \bibnamefont
  {Speagle}},\ }\href@noop {} {\bibfield  {journal} {\bibinfo  {journal}
  {{Monthly Notices of the Royal Astronomical Society}}\ } (\bibinfo {year}
  {2020})},\ \Eprint {http://arxiv.org/abs/1904.02180} {arXiv:1904.02180
  [astro-ph]} \BibitemShut {NoStop}%
\bibitem [{\citenamefont {Collaboration}\ and\ \citenamefont
  {Collaboration}(2022)}]{ligo_scientific_collaboration_and_virgo_2022_6513631}%
  \BibitemOpen
  \bibfield  {author} {\bibinfo {author} {\bibfnamefont {L.~S.}\ \bibnamefont
  {Collaboration}}\ and\ \bibinfo {author} {\bibfnamefont {V.}~\bibnamefont
  {Collaboration}},\ }\href {\doibase 10.5281/zenodo.6513631} {} (\bibinfo
  {year} {2022})\BibitemShut {NoStop}%
\bibitem [{\citenamefont {Collaboration}\ \emph {et~al.}(2021)\citenamefont
  {Collaboration}, \citenamefont {Collaboration},\ and\ \citenamefont
  {Collaboration}}]{ligo_scientific_collaboration_and_virgo_2021_5546663}%
  \BibitemOpen
  \bibfield  {author} {\bibinfo {author} {\bibfnamefont {L.~S.}\ \bibnamefont
  {Collaboration}}, \bibinfo {author} {\bibfnamefont {V.}~\bibnamefont
  {Collaboration}}, \ and\ \bibinfo {author} {\bibfnamefont {K.}~\bibnamefont
  {Collaboration}},\ }\href {\doibase 10.5281/zenodo.5546663} {} (\bibinfo
  {year} {2021})\BibitemShut {NoStop}%
\bibitem [{\citenamefont {Rodriguez}\ \emph {et~al.}(2013)\citenamefont
  {Rodriguez}, \citenamefont {Farr}, \citenamefont {Farr},\ and\ \citenamefont
  {Mandel}}]{Rodriguez:2013mla}%
  \BibitemOpen
  \bibfield  {author} {\bibinfo {author} {\bibfnamefont {C.~L.}\ \bibnamefont
  {Rodriguez}}, \bibinfo {author} {\bibfnamefont {B.}~\bibnamefont {Farr}},
  \bibinfo {author} {\bibfnamefont {W.~M.}\ \bibnamefont {Farr}}, \ and\
  \bibinfo {author} {\bibfnamefont {I.}~\bibnamefont {Mandel}},\ }\href
  {\doibase 10.1103/PhysRevD.88.084013} {\bibfield  {journal} {\bibinfo
  {journal} {Phys. Rev. D}\ }\textbf {\bibinfo {volume} {88}},\ \bibinfo
  {pages} {084013} (\bibinfo {year} {2013})},\ \Eprint
  {http://arxiv.org/abs/1308.1397} {arXiv:1308.1397 [astro-ph.IM]} \BibitemShut
  {NoStop}%
\bibitem [{\citenamefont {Mandel}\ \emph {et~al.}(2014)\citenamefont {Mandel},
  \citenamefont {Berry}, \citenamefont {Ohme}, \citenamefont {Fairhurst},\ and\
  \citenamefont {Farr}}]{Mandel:2014tca}%
  \BibitemOpen
  \bibfield  {author} {\bibinfo {author} {\bibfnamefont {I.}~\bibnamefont
  {Mandel}}, \bibinfo {author} {\bibfnamefont {C.~P.~L.}\ \bibnamefont
  {Berry}}, \bibinfo {author} {\bibfnamefont {F.}~\bibnamefont {Ohme}},
  \bibinfo {author} {\bibfnamefont {S.}~\bibnamefont {Fairhurst}}, \ and\
  \bibinfo {author} {\bibfnamefont {W.~M.}\ \bibnamefont {Farr}},\ }\href
  {\doibase 10.1088/0264-9381/31/15/155005} {\bibfield  {journal} {\bibinfo
  {journal} {Class. Quant. Grav.}\ }\textbf {\bibinfo {volume} {31}},\ \bibinfo
  {pages} {155005} (\bibinfo {year} {2014})},\ \Eprint
  {http://arxiv.org/abs/1404.2382} {arXiv:1404.2382 [gr-qc]} \BibitemShut
  {NoStop}%
\bibitem [{\citenamefont {Poisson}(1995)}]{Poisson:1995vs}%
  \BibitemOpen
  \bibfield  {author} {\bibinfo {author} {\bibfnamefont {E.}~\bibnamefont
  {Poisson}},\ }\href {\doibase 10.1103/PhysRevD.52.5719} {\bibfield  {journal}
  {\bibinfo  {journal} {Phys. Rev. D}\ }\textbf {\bibinfo {volume} {52}},\
  \bibinfo {pages} {5719} (\bibinfo {year} {1995})},\ \bibinfo {note}
  {[Addendum: Phys.Rev.D 55, 7980--7981 (1997)]},\ \Eprint
  {http://arxiv.org/abs/gr-qc/9505030} {arXiv:gr-qc/9505030} \BibitemShut
  {NoStop}%
\bibitem [{\citenamefont {Moore}\ \emph {et~al.}(2018)\citenamefont {Moore},
  \citenamefont {Robson}, \citenamefont {Loutrel},\ and\ \citenamefont
  {Yunes}}]{Moore:2018kvz}%
  \BibitemOpen
  \bibfield  {author} {\bibinfo {author} {\bibfnamefont {B.}~\bibnamefont
  {Moore}}, \bibinfo {author} {\bibfnamefont {T.}~\bibnamefont {Robson}},
  \bibinfo {author} {\bibfnamefont {N.}~\bibnamefont {Loutrel}}, \ and\
  \bibinfo {author} {\bibfnamefont {N.}~\bibnamefont {Yunes}},\ }\href
  {\doibase 10.1088/1361-6382/aaea00} {\bibfield  {journal} {\bibinfo
  {journal} {Class. Quant. Grav.}\ }\textbf {\bibinfo {volume} {35}},\ \bibinfo
  {pages} {235006} (\bibinfo {year} {2018})},\ \Eprint
  {http://arxiv.org/abs/1807.07163} {arXiv:1807.07163 [gr-qc]} \BibitemShut
  {NoStop}%
\bibitem [{\citenamefont {Chatziioannou}\ \emph {et~al.}(2017)\citenamefont
  {Chatziioannou}, \citenamefont {Klein}, \citenamefont {Yunes},\ and\
  \citenamefont {Cornish}}]{Chatziioannou:2017tdw}%
  \BibitemOpen
  \bibfield  {author} {\bibinfo {author} {\bibfnamefont {K.}~\bibnamefont
  {Chatziioannou}}, \bibinfo {author} {\bibfnamefont {A.}~\bibnamefont
  {Klein}}, \bibinfo {author} {\bibfnamefont {N.}~\bibnamefont {Yunes}}, \ and\
  \bibinfo {author} {\bibfnamefont {N.}~\bibnamefont {Cornish}},\ }\href
  {\doibase 10.1103/PhysRevD.95.104004} {\bibfield  {journal} {\bibinfo
  {journal} {Phys. Rev. D}\ }\textbf {\bibinfo {volume} {95}},\ \bibinfo
  {pages} {104004} (\bibinfo {year} {2017})},\ \Eprint
  {http://arxiv.org/abs/1703.03967} {arXiv:1703.03967 [gr-qc]} \BibitemShut
  {NoStop}%
\bibitem [{\citenamefont {Moore}\ \emph {et~al.}(2016)\citenamefont {Moore},
  \citenamefont {Berry}, \citenamefont {Chua},\ and\ \citenamefont
  {Gair}}]{Moore:2015sza}%
  \BibitemOpen
  \bibfield  {author} {\bibinfo {author} {\bibfnamefont {C.~J.}\ \bibnamefont
  {Moore}}, \bibinfo {author} {\bibfnamefont {C.~P.~L.}\ \bibnamefont {Berry}},
  \bibinfo {author} {\bibfnamefont {A.~J.~K.}\ \bibnamefont {Chua}}, \ and\
  \bibinfo {author} {\bibfnamefont {J.~R.}\ \bibnamefont {Gair}},\ }\href
  {\doibase 10.1103/PhysRevD.93.064001} {\bibfield  {journal} {\bibinfo
  {journal} {Phys. Rev. D}\ }\textbf {\bibinfo {volume} {93}},\ \bibinfo
  {pages} {064001} (\bibinfo {year} {2016})},\ \Eprint
  {http://arxiv.org/abs/1509.04066} {arXiv:1509.04066 [gr-qc]} \BibitemShut
  {NoStop}%
\bibitem [{\citenamefont {Doctor}\ \emph {et~al.}(2017)\citenamefont {Doctor},
  \citenamefont {Farr}, \citenamefont {Holz},\ and\ \citenamefont
  {P\"urrer}}]{Doctor:2017csx}%
  \BibitemOpen
  \bibfield  {author} {\bibinfo {author} {\bibfnamefont {Z.}~\bibnamefont
  {Doctor}}, \bibinfo {author} {\bibfnamefont {B.}~\bibnamefont {Farr}},
  \bibinfo {author} {\bibfnamefont {D.~E.}\ \bibnamefont {Holz}}, \ and\
  \bibinfo {author} {\bibfnamefont {M.}~\bibnamefont {P\"urrer}},\ }\href
  {\doibase 10.1103/PhysRevD.96.123011} {\bibfield  {journal} {\bibinfo
  {journal} {Phys. Rev. D}\ }\textbf {\bibinfo {volume} {96}},\ \bibinfo
  {pages} {123011} (\bibinfo {year} {2017})},\ \Eprint
  {http://arxiv.org/abs/1706.05408} {arXiv:1706.05408 [astro-ph.HE]}
  \BibitemShut {NoStop}%
\bibitem [{\citenamefont {Edelman}\ \emph {et~al.}(2021)\citenamefont {Edelman}
  \emph {et~al.}}]{Edelman:2020aqj}%
  \BibitemOpen
  \bibfield  {author} {\bibinfo {author} {\bibfnamefont {B.}~\bibnamefont
  {Edelman}} \emph {et~al.},\ }\href {\doibase 10.1103/PhysRevD.103.042004}
  {\bibfield  {journal} {\bibinfo  {journal} {Phys. Rev. D}\ }\textbf {\bibinfo
  {volume} {103}},\ \bibinfo {pages} {042004} (\bibinfo {year} {2021})},\
  \Eprint {http://arxiv.org/abs/2008.06436} {arXiv:2008.06436 [gr-qc]}
  \BibitemShut {NoStop}%
\bibitem [{\citenamefont {Jan}\ \emph {et~al.}(2020)\citenamefont {Jan},
  \citenamefont {Yelikar}, \citenamefont {Lange},\ and\ \citenamefont
  {O'Shaughnessy}}]{Jan:2020bdz}%
  \BibitemOpen
  \bibfield  {author} {\bibinfo {author} {\bibfnamefont {A.~Z.}\ \bibnamefont
  {Jan}}, \bibinfo {author} {\bibfnamefont {A.~B.}\ \bibnamefont {Yelikar}},
  \bibinfo {author} {\bibfnamefont {J.}~\bibnamefont {Lange}}, \ and\ \bibinfo
  {author} {\bibfnamefont {R.}~\bibnamefont {O'Shaughnessy}},\ }\href {\doibase
  10.1103/PhysRevD.102.124069} {\bibfield  {journal} {\bibinfo  {journal}
  {Phys. Rev. D}\ }\textbf {\bibinfo {volume} {102}},\ \bibinfo {pages}
  {124069} (\bibinfo {year} {2020})},\ \Eprint
  {http://arxiv.org/abs/2011.03571} {arXiv:2011.03571 [gr-qc]} \BibitemShut
  {NoStop}%
\bibitem [{\citenamefont {Hu}\ and\ \citenamefont {Veitch}(2022)}]{Hu:2022rjq}%
  \BibitemOpen
  \bibfield  {author} {\bibinfo {author} {\bibfnamefont {Q.}~\bibnamefont
  {Hu}}\ and\ \bibinfo {author} {\bibfnamefont {J.}~\bibnamefont {Veitch}},\
  }\href {\doibase 10.1103/PhysRevD.106.044042} {\bibfield  {journal} {\bibinfo
   {journal} {Phys. Rev. D}\ }\textbf {\bibinfo {volume} {106}},\ \bibinfo
  {pages} {044042} (\bibinfo {year} {2022})},\ \Eprint
  {http://arxiv.org/abs/2205.08448} {arXiv:2205.08448 [gr-qc]} \BibitemShut
  {NoStop}%
\bibitem [{\citenamefont {Read}(2023)}]{Read:2023hkv}%
  \BibitemOpen
  \bibfield  {author} {\bibinfo {author} {\bibfnamefont {J.~S.}\ \bibnamefont
  {Read}},\ }\href@noop {} {\  (\bibinfo {year} {2023})},\ \Eprint
  {http://arxiv.org/abs/2301.06630} {arXiv:2301.06630 [gr-qc]} \BibitemShut
  {NoStop}%
\bibitem [{\citenamefont {Yagi}\ and\ \citenamefont
  {Yunes}(2013{\natexlab{a}})}]{Yagi:2013bca}%
  \BibitemOpen
  \bibfield  {author} {\bibinfo {author} {\bibfnamefont {K.}~\bibnamefont
  {Yagi}}\ and\ \bibinfo {author} {\bibfnamefont {N.}~\bibnamefont {Yunes}},\
  }\href {\doibase 10.1126/science.1236462} {\bibfield  {journal} {\bibinfo
  {journal} {Science}\ }\textbf {\bibinfo {volume} {341}},\ \bibinfo {pages}
  {365} (\bibinfo {year} {2013}{\natexlab{a}})},\ \Eprint
  {http://arxiv.org/abs/1302.4499} {arXiv:1302.4499 [gr-qc]} \BibitemShut
  {NoStop}%
\bibitem [{\citenamefont {Yagi}\ and\ \citenamefont
  {Yunes}(2013{\natexlab{b}})}]{Yagi:2013awa}%
  \BibitemOpen
  \bibfield  {author} {\bibinfo {author} {\bibfnamefont {K.}~\bibnamefont
  {Yagi}}\ and\ \bibinfo {author} {\bibfnamefont {N.}~\bibnamefont {Yunes}},\
  }\href {\doibase 10.1103/PhysRevD.88.023009} {\bibfield  {journal} {\bibinfo
  {journal} {Phys. Rev. D}\ }\textbf {\bibinfo {volume} {88}},\ \bibinfo
  {pages} {023009} (\bibinfo {year} {2013}{\natexlab{b}})},\ \Eprint
  {http://arxiv.org/abs/1303.1528} {arXiv:1303.1528 [gr-qc]} \BibitemShut
  {NoStop}%
\bibitem [{\citenamefont {Chatziioannou}(2022)}]{Chatziioannou:2021tdi}%
  \BibitemOpen
  \bibfield  {author} {\bibinfo {author} {\bibfnamefont {K.}~\bibnamefont
  {Chatziioannou}},\ }\href {\doibase 10.1103/PhysRevD.105.084021} {\bibfield
  {journal} {\bibinfo  {journal} {Phys. Rev. D}\ }\textbf {\bibinfo {volume}
  {105}},\ \bibinfo {pages} {084021} (\bibinfo {year} {2022})},\ \Eprint
  {http://arxiv.org/abs/2108.12368} {arXiv:2108.12368 [gr-qc]} \BibitemShut
  {NoStop}%
\bibitem [{\citenamefont {Abbott}\ \emph
  {et~al.}(2018{\natexlab{b}})\citenamefont {Abbott} \emph
  {et~al.}}]{LIGOScientific:2018cki}%
  \BibitemOpen
  \bibfield  {author} {\bibinfo {author} {\bibfnamefont {B.~P.}\ \bibnamefont
  {Abbott}} \emph {et~al.} (\bibinfo {collaboration} {LIGO Scientific,
  Virgo}),\ }\href {\doibase 10.1103/PhysRevLett.121.161101} {\bibfield
  {journal} {\bibinfo  {journal} {Phys. Rev. Lett.}\ }\textbf {\bibinfo
  {volume} {121}},\ \bibinfo {pages} {161101} (\bibinfo {year}
  {2018}{\natexlab{b}})},\ \Eprint {http://arxiv.org/abs/1805.11581}
  {arXiv:1805.11581 [gr-qc]} \BibitemShut {NoStop}%
\bibitem [{\citenamefont {Perkins}\ and\ \citenamefont
  {Yunes}(2022)}]{Perkins:2022fhr}%
  \BibitemOpen
  \bibfield  {author} {\bibinfo {author} {\bibfnamefont {S.}~\bibnamefont
  {Perkins}}\ and\ \bibinfo {author} {\bibfnamefont {N.}~\bibnamefont
  {Yunes}},\ }\href {\doibase 10.1103/PhysRevD.105.124047} {\bibfield
  {journal} {\bibinfo  {journal} {Phys. Rev. D}\ }\textbf {\bibinfo {volume}
  {105}},\ \bibinfo {pages} {124047} (\bibinfo {year} {2022})},\ \Eprint
  {http://arxiv.org/abs/2201.02542} {arXiv:2201.02542 [gr-qc]} \BibitemShut
  {NoStop}%
\end{thebibliography}%
\end{document}